# Evaluating Computational Shortcuts in Supercell-Based Phonon Calculations of Molecular Crystals: The Instructive Case of Naphthalene

*Tomas Kamencek[1,2], Sandro Wieser[1], Hirotaka Kojima[3], Natalia Bedoya-Martínez[4], Johannes P. Dürholt[5], Rochus Schmid[5], and Egbert Zojer[1]*

[1]*Institute of Solid State Physics, Graz University of Technology, NAWI Graz, Petersgasse 16, 8010 Graz, Austria*

[2]*Institute of Physical and Theoretical Chemistry, Graz University of Technology, NAWI Graz, Stremayrgasse 9, 8010 Graz, Austria*

[3]*Division of Materials Science, Nara Institute of Science and Technology, 8916-5 Takayama, Ikoma, Nara 630-0192, Japan*

[4]*Materials Center Leoben, Roseggerstraße 12, 8700 Leoben, Austria*

[5]*Computational Materials Chemistry Group, Faculty of Chemistry and Biochemistry, Ruhr-University Bochum, Universitätsstraße 150, 44801 Bochum, Germany*

**ABSTRACT:** Phonons crucially impact a variety of properties of organic semiconductor materials. For instance, charge- and heat transport depend on low-frequency phonons, while for other properties, such as the free energy, especially high-frequency phonons count. For all these quantities one needs to know the entire phonon band structure, whose simulation becomes exceedingly expensive for more complex systems when using methods like dispersion-corrected density functional theory (DFT). Therefore, in the present contribution we evaluate the performance of more approximate methodologies, including density functional tight binding (DFTB) and a pool of force fields (FF) of varying complexity and sophistication. Beyond merely comparing phonon band structures, we also critically evaluate to what extent derived quantities, like temperature-dependent heat capacities, mean squared thermal displacements and temperature-dependent free energies are impacted by shortcomings in the description of the phonon bands. As a benchmark system, we choose (deuterated) naphthalene, as the only organic semiconductor material for which to date experimental phonon band structures are available in the literature. Overall, the best performance amongst the approximate methodologies is observed for a system-specifically parametrized second-generation force field. Interestingly, in the low-frequency regime also force fields with a rather simplistic model for the bonding interactions (like the General Amber Force Field) perform rather well. As far as the tested DFTB parametrization is concerned, we obtain a significant underestimation of the unit cell volume resulting in a pronounced overestimation of the phonon energies in the low frequency region. This cannot be mended by relying on the DFT-calculated unit cell, since with this unit cell the DFTB phonon frequencies significantly underestimate the experiments.

# 1. INTRODUCTION

Over the past years, molecular crystals have been the subject of numerous studies aiming at a better understanding of their properties in order to improve their performance in organic-semiconductor based devices. Many of these properties are crucially influenced by phonons. For example, a strong electron-phonon coupling is one of the main factors hampering charge transport in organic semiconductors[1–6]. Phonons are also important for other dynamic processes like thermal transport[7] or thermoelectricity[8,9]. Furthermore, the phonon contribution to the free energy is often found to be crucial for correctly predicting the relative stability of different phases[10–12], especially when dealing with polymorphs that are very close in energy, as it is often the case in molecular crystals[13,14]. Consequently, a detailed knowledge of the phonon band structure is highly beneficial.

Unfortunately, for materials as complex as molecular crystals, it is difficult to reliably determine phonon bands both in experiments and in simulations: inelastic neutron scattering, which is typically used to measure phonon band structures, requires large single crystals, which are often difficult to grow. Moreover, these crystals ought to consist of deuterated molecules, as this results in higher coherent and lower incoherent scattering cross-sections for neutrons[15–17]. As a consequence, to the best of our knowledge, the only crystal consisting of π-conjugated molecules for which experimental phonon band structure data are available is deuterated naphthalene[18].

Complex molecular crystals relevant in organic devices also pose a significant challenge for simulations. To date, the most common approach is to calculate phonon bands from the dynamical matrix of the system either by density-functional perturbation theory[19,20] or by finite displacements in supercells. Since the current work aims at a quantitative comparison of different low-level theories, here the method of choice for calculating phonon band structures is the finite displacement approach in order to consistently apply the same methodology throughout all levels of theory. Within this approach, the dynamical matrix is derived from the forces on each atom in the unit cell caused by cartesian displacements of all other atoms contained in rather large supercells. Such calculations are significantly more costly than, for example, the simulation of infrared (IR) or Raman spectra of the same molecular crystal, as in that case only Γ-point phonons are relevant, and simulations can be restricted to primitive unit cells.

For calculating Γ-point phonons, density functional theory (DFT)[21,22] is typically the method of choice. Indeed, as shown recently for a variety of conjugated materials and their polymorphs,[23] combining DFT with a suitably chosen van der Waals correction, yields an excellent agreement with experimental Raman data even in the range between ~5 cm$^{-1}$ and 100 cm$^{-1}$. As indicated above, the situation becomes computationally more challenging, as soon as phonons in the entire first Brillouin zone (1BZ) need to be considered. This is, for example, the case for electron scattering processes, thermal transport, or the reliable prediction of the phase stability. Nevertheless, free energies are often calculated with Γ-phonons only[24,25], because the entire 1BZ is hardly accessible with *ab initio* methods like dispersion-corrected DFT. This can, however, lead to severe discrepancies in thermodynamic quantities as discussed in the Supporting Information.

A possible strategy for reducing the computational cost is to resort to lower levels of theory for describing interatomic and intermolecular interactions. This raises the questions, under which circumstances such approaches could be used, and how accurate the calculated phonon dispersion must be to obtain reliable phonon related properties (such as thermodynamic potentials, group velocities, thermal displacement etc.). Addressing these questions is at the very heart of the present paper, where the phonon properties of (deuterated) naphthalene are analyzed in the framework of different levels of theory. The latter comprise dispersion corrected density functional theory, density functional tight binding (DFTB)[26–29] and various flavors of classical force fields (FF).

The paper is organized as follows: after a brief description of the studied system, the importance of evaluating the reliability of the different methods separately in the low-frequency regime and for the entire spectral range is discussed. Subsequently, the reliability of the DFT data as a reference for the further analysis is shown. This is done by comparing the simulated DFT phonon band structure of deuterated naphthalene to the available experimental data[18]. As part of this comparison, different a posteriori van-der-Waals (vdW) corrections within DFT are tested, extending a previous study by Brown-Altvater et al.[30]. DFTB- and FF-calculated phonon bands are, subsequently, compared to the results of the aforementioned DFT reference results. Finally, as a key aspect of this manuscript, the impact of the differences in the calculated band structures on the derived practically relevant phonon properties is evaluated. The latter comprise thermal atomic motion (i.e. mean squared thermal displacements), heat capacities, group velocities, and thermodynamic potentials.

## 1.1 The studied system

Due to the availability of suitable experimental data (see above), we will focus on crystalline (deuterated) naphthalene. The crystal consists of two molecules per unit cell with 18 atoms each. Therefore, the unit cell has 108 (=36×3) degrees of freedom (i.e. phonon bands): twelve of these are dominated by intermolecular (three rotational and three translational degrees of freedom per molecule), and 96 by intramolecular motions. The system crystallizes in a monoclinic Bravais lattice with space group $P2_1/a$. Two orthographic projections of the unit cell are shown in Figure 1. The naphthalene molecules arrange in a herringbone fashion in 2D layers, with the lattice vector *b* being much shorter than the remaining two (|*a*|=8.08 Å, |*b*|=5.93 Å, |*c*|=8.63 Å). This suggests anisotropic (phonon) properties.

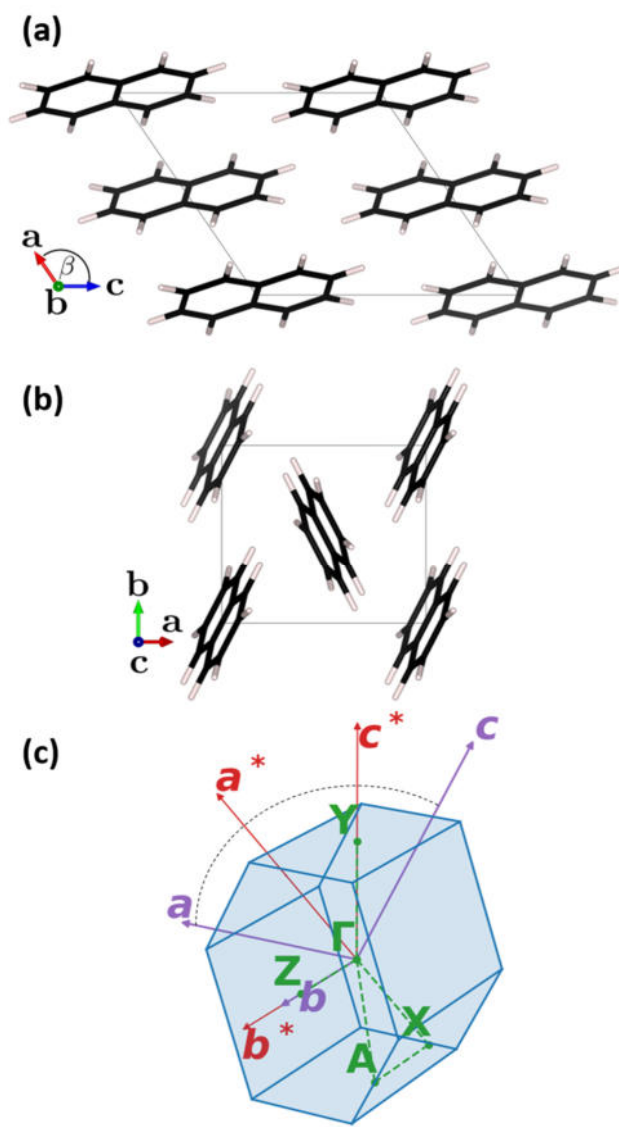

***Figure 1.*** *Unit cell of naphthalene seen (a) along lattice vector **b**, and (b) along lattice vector **c**. Visualized with VESTA 3[31]. Panel (c) shows the first Brillouin zone of the crystal, where the primitive lattice vectors (a,b,c) and as the reciprocal primitive lattice vectors (**a\***,**b\***,**c\***) are indicated by purple and red arrows, respectively. The vectors **a**, **a\***, **c**, and **c\*** lie in one plane (indicated by the black dashed arc) perpendicular to **b** and **b\***. The path connecting high symmetry points used in band diagrams is shown as green dashed lines.*

Figure 1(c) shows the first Brillouin zone (1BZ) of the crystal and the high-symmetry points used in the band diagrams throughout this paper. Note that **b** and its reciprocal lattice vector **b*** are collinear, while all other real space and reciprocal lattice vectors (**a**, **a***, **c**, and **c***) lie in the same plane, which is perpendicular to **b** and **b***. Thus, The ΓZ direction in the band structures corresponds to the direction along the shortest lattice vector, **b**. ΓY and ΓX are directions slightly inclined with respect to the long and short molecular axis, respectively.

Finally, it should be noted that with one exception, in the following discussion we will focus on ordinary, protonated naphthalene crystals. This exception is the comparison of DFT-calculated phonon band structures with neutron scattering experiments. There, we will discuss simulations on a fully deuterated system (i.e., we changed the mass of the hydrogen atoms in the phonon calculations) in order to be consistent with the experimental situation.

## 2. METHODOLOGY

### 2.1 Density Functional Theory calculations

DFT calculations were carried out with the *VASP* code[32–35] (version 5.4.1), using the PBE functional[36] and employing the recommended standard potentials[37] within the projector-augmented wave method[38]. The occupation of electronic states was described with a Gaussian smearing of $\sigma = 0.05$ eV. For calculations of primitive unit cells, the electronic band structure was sampled with a Γ-centered 2×3×2 k-mesh, while for supercell calculations (2×3×2 super cells, see below) only electronic states at the Γ-point were considered. This choice is based on convergence tests for the Γ-point phonons when varying the sampling of the 1BZ in the electronic structure calculations (see Supporting Information). A plane wave energy cutoff of 900 eV, a SCF energy convergence criterion of $10^{-8}$ eV and the global precision `Accurate` were used (for details see

VASP manual[39]). These parameters ensure DFT reference calculations with a high accuracy. A more detailed description of how the specific settings impact the computational time and the accuracy of the results can be found in the Supporting Information. The atomic positions and the lattice parameters were optimized to residual forces below 0.5 meV/Å employing the conjugate gradient algorithm. The final lattice geometry was found by fitting structures optimized with fixed unit-cell volume to a Vinet equation of state[40]. To account for van der Waals (vdW) interactions, by default the D3 correction with Becke-Johnson damping (D3-BJ) was used after careful tests (see Section 3.2.1; the used standard parameters are listed in the Supporting Information). This approach employs a $r^{-6}$ and a $r^{-8}$ term to describe attractive vdW interaction with the coefficients being dependent on the chemical environment of each atom by means of geometry-dependent coordination numbers[41,42].

The Raman activities for isolated molecules were calculated using the *Gaussian 16* package (Revision A.03)[43], while the Raman activities (equations based on ref 44) of the crystalline phase were calculated by finite (cartesian) displacements with our own post-processing tool, as described in detail in the Supporting Information.

## 2.2 Density Functional Tight Binding calculations

The *DFTB+* package[45] (version 18.1) was employed to carry out the calculations within the self-consistent charge (SCC-)DFTB approach. The publicly available *3ob-3-1* Slater-Koster files including the *3ob:freq-1-2* extension for obtaining more accurate vibrational properties[46] were used throughout all calculations. The third order correction of the DFTB3 functional[47] was included, as the used functional-dependent vdW parameters were optimized for this functional. The D3-BJ correction was used to be consistent with the DFT calculations employing (standard) parameters listed in the Supporting Information. Regarding the sampling of reciprocal space, the same settings as for DFT were chosen. All available angular momentum atomic orbitals for each species were considered, and the SCC convergence criterion was set to $10^{-10}$ elementary charges.

The optimization of the primitive unit cell shapes turned out to be more involved than in the DFT calculations. In order to keep the monoclinic Bravais lattice (independent variables are the lengths of the unit cell vectors, |*a*|, |*b*|, and |*c*|, and the angle β≠90°), several constraints would have been required during the optimization. These are not implemented in the used version of the code. In order to overcome this problem, the lengths of the lattice vectors were optimized together with the atomic positions for a set of fixed monoclinic angles β, using the conjugate gradient algorithm. To find the optimal monoclinic angle, a second order polynomial was then fitted to the energy-vs.-β curves to find the optimal monoclinic angle (see Supporting Information). The resulting angle was subsequently used for a final optimization of the lengths of the lattice vectors and the atomic positions (ensuring residual forces below $10^{-8}$ eV/Å).

In this context it is interesting to mention that Brandenburg and Grimme suggested using DFT optimized unit-cell parameters when calculating phonon properties employing DFTB[48]. The suitability of this approach for the present problem will also be tested.

Finally, we want to emphasize that our results within DFTB have been obtained in an "off-the-shelf" manner – i.e. we did not reparametrize any Slater-Koster files, but rather relied on the publicly available ones.

## 2.3 Force field calculations

The performance of empirical force fields (FFs) at various levels of sophistication was also assessed. As a starting point, we employed the Generalized AMBER Force Field (GAFF)[49], where AMBER stands for "Assisted Model Building with Energy Refinement". It is a transferable force-field frequently used for simulations of organic semiconductors[7,50–52]. The GAFF parametrization has been specifically designed for small organic molecules. In GAFF, all bonded interactions are described by harmonic potentials and no cross terms between different geometric parameters are considered. Electrostatic interactions are described via individual point charges localized at the positions of the nuclei. As GAFF provides no predefined atomic charges, they were determined here from the electrostatic potential of the periodic DFT reference (with the DFT-optimized unit cell and atomic positions) employing the REPEAT[53] method.

The description of interatomic interactions in GAFF is only harmonic – i.e. all non-parabolic potential terms arise from Coulomb and vdW interaction. Therefore, we also tested the performance of a more sophisticated, second-generation force field building on the *"condensed-phase optimized molecular potentials for atomistic simulation studies"* (COMPASS)[54], which has been used in several studies dealing with transport properties of molecular crystals[55–58]. The parameters in COMPASS have been specifically developed for aliphatic and aromatic compounds. In contrast to GAFF, COMPASS includes anharmonic bonded interactions and numerous cross-terms between bonds, bending angles and torsions. The inclusion of these terms should lead to significant improvements for vibrational properties. A further difference is the softer 9-6 Lennard-Jones potential in COMPASS (see refs 59,60) compared to the 12-6 term utilized in GAFF[49]. Notably, in the COMPASS simulations, standard pre-defined atomic charges were used.

It should be stressed that all methods discussed so far rely on off-the-shelf implementations – either using standard PAW potentials in DFT, publicly available Slater-Koster files in DFTB, or ready-to-use force fields. This does not (directly) apply to our final option, a non-transferable FF, which has been parametrized based on DFT reference data of the studied system. The functional form of that force field is based on the MOF-FF[61] class of force fields, which has originally been developed for metal-organic frameworks (MOFs)[62]. In fact, to the best of our knowledge, MOF-FF type force fields have not been applied to molecular crystals in the past. Our parametrization of MOF-FF contains the original cross terms from the implementation in ref 61. In analogy to the COMPASS FF, for atoms bonded in a 1-2-3-4 fashion we also considered cross terms between stretching motions of atoms 1 and 2 as well as between atoms 3 and 4 (so-called *bb13* cross terms; see Supporting Information). Another deviation is the different description of vdW interactions: while COMPASS uses the 9-6 Lennard-Jones potential, in MOF-FF a damped Buckingham potential is used – i.e., MOF-FF relies on an ordinary Buckingham potential (consisting of an exponential repulsive term and an attractive $r^{-6}$ term), where the attractive part is additionally multiplied with a damping function. The latter eliminates the potential minimum of the undamped version at small distances (for a thorough definition see ref 61). Additionally, MOF-FF employs spherical Gaussian charge distributions instead of point charges to describe electrostatic interactions. Finally, the number of atom types, for which distinct parameters are considered, differs in the above-mentioned force fields. For example, for the naphthalene molecules, MOF-FF distinguishes between five different atom types (two hydrogens and three carbons), as opposed to only two (one hydrogen and one carbon) considered in GAFF or COMPASS. The parameterization of MOF-FF was performed by using the software FFgen[61,63]. Our fit is based on molecular ab-initio reference data comprising interatomic force constants (i.e., the Hessian matrix) and the optimized geometry (bond lengths, angles, dihedrals, etc.) for an isolated naphthalene molecule obtained by the Turbomole[64] software package (version 7.3). Specific simulation details can be found in the Supporting Information together with the fitted force-field parameters. Atomic charges were obtained from a fit of the electrostatic potential of the isolated naphthalene molecule in the gas phase using the Horton[65] package (see Supporting Information). The vdW parameters used within the MOF-FF are derived from the MM3 parameter set[66,67] with certain modifications described in detail in ref 61.

The geometry optimizations and the calculations of interatomic forces for all FFs were performed with the *LAMMPS*[68] package. The geometries were minimized to residual forces below $10^{-7}$ eV/Å with the conjugate gradient algorithm. A cutoff for both, vdW and Coulomb interaction, of 12 Å was chosen. To avoid discontinuities at this cutoff, an additional smoothening between 10.8 and 12 Å was applied for MOF-FF and GAFF. For COMPASS we did not apply such a procedure, as such a smoothening is not available without changing the force field.

It should be stressed that in the present paper we are concerned with phonon band structures. Therefore, in contrast to the more typical applications of FFs in molecular dynamics simulations at finite temperatures, we restricted our simulations to lattice dynamics performed at 0 K. Here, the force fields are solely employed as a means for obtaining interatomic forces in analogy to our quantum-mechanical simulations.

## 2.4 Phonon calculations

The PHONOPY[69] code was used to calculate phonon bands by means of finite displacements in supercells. In the case of DFT and DFTB, 2×3×2 supercells were found to be large enough for obtaining a converged dynamical matrix. For the FF-based calculations, 3×3×3 supercells were necessary to reach the desired level of accuracy (see Supporting Information). The default PHONOPY displacement of 0.01 Å was used for all calculations, as varying the displacement amplitudes between 0.0025 Å and 0.02 Å resulted in maximum absolute frequency differences below 0.02 THz for DFT. The obtained harmonic force constants were symmetrized a-posteriori with PHONOPY's internal subroutines, in order to correct for possibly lost symmetries caused by numerical inaccuracies. Group velocities were calculated from the analytic gradients of the dynamical matrices with PHONOPY.

For plotting phonon bands, the reciprocal space was sampled at 200 wave vectors, $q$, between each pair of high-symmetry points in the 1BZ. For comparing quantities in the entire 1BZ, or for quantities for which a Brillouin zone integration (summation) is required, 9×10×9 $q$-meshes were used to sample the different directions in the anisotropic unit cell as uniformly as possible. This choice of $q$-mesh yields a $q$-step size of ~0.1 Å$^{-1}$ in each direction with 810 $q$-points per band (246 irreducible ones for the given space group symmetry). For several of the comparisons below, phonon properties (frequencies or group velocities) are plotted at these discrete $q$-points. For group velocities and thermal displacements (see below), the $q$-meshes have been shifted such that they do not include the Γ point. In the case of group velocities, this is necessary to obtain unbiased estimates with respect to the reference because at Γ there is not a single uniquely defined group velocity for the acoustic phonons. This would result in incorrect errors if Γ-phonons were included. For the thermal displacements, shifted meshes have been used to avoid divergences at zero frequency (see below).

Smooth curves, like for the densities of states (DOS), are obtained by summing over Lorentzian functions with widths σ of 0.05 THz (2σ corresponds to the full width at half maximum; FWHM) centered at the frequencies of the phonons calculated employing the above-described $q$-mesh. The resulting DOSs are then normalized such that their integral over frequency yields 3$N$, where $N$ is the number of atoms in the primitive unit cell. Thermodynamic quantities were calculated according to the well-known expressions from statistical physics (see below). It should be stressed that the reported temperature dependence of those thermodynamic properties does not account for thermal expansion of the lattice since such anharmonic effects lie beyond the scope of the current work.

## 3. RESULTS AND DISCUSSION

### 3.1 Relevant frequency ranges

In the following discussion, we will separately benchmark the performance of the different methodologies for the low-frequency region (up to frequencies of 9 THz, corresponding to ~300 cm$^{-1}$, see below) and for the entire spectral range in which vibrations occur. One of the motivations for distinguishing between frequency ranges is that several relevant quantities are primarily impacted by phonons at rather low frequencies.

This, for example, applies to the mean squared thermal displacement (MSTD) $\langle|u_i^\alpha|^2\rangle$ of atom $\alpha$ in direction $i$ (see eq 1a) with $m_\alpha$ being the mass of that atom.

$$\langle|u_i^\alpha|^2\rangle = \frac{\hbar}{2m_\alpha} \int d\omega \, f_D(\omega, T) \, |e_i^\alpha|^2 \, DOS(\omega) \qquad (1a)$$

$$f_D(\omega, T) = \frac{1 + 2n(\omega, T)}{\omega} \qquad (1b)$$

This quantity is a measure for thermal disorder, which, for example, crucially impacts the electrical conductivity of organic semiconductors. Furthermore, the (anisotropic) mean square thermal displacements play a significant role in X-ray diffraction (XRD), as they enter the Debye-Waller factors[70,71], in this way determining the width of XRD peaks due to atomic motion.

The $\langle|u_i^\alpha|^2\rangle$ can be decomposed into material-specific quantities, like the absolute values of the phonon eigenvectors (polarization vectors $e_i^\alpha$) and the density of states (DOS), as well as into a material-independent function $f_D$ (see eq 1b, which scales the contributions of the individual phonon modes[72]. $f_D$ is determined by the mode occupation $n(\omega,T)$ given by the Bose-Einstein distribution and by the angular frequency ω of the mode. As shown in Figure 2(a), $f_D$ exclusively "selects" low-frequency modes and at low temperatures converges to the hyperbolic $\omega^{-1}$ function.

Similarly, the phonon contribution to the heat capacity (see eq 2a) can be calculated as an integral over the DOS multiplied with a material-independent spectral function $f_C$ (given in eq 2b).

$$C_V = k_B \int d\omega \, DOS(\omega) f_C(\omega, T) \qquad (2a)$$

$$f_C(\omega, T) = k_B \frac{(x/2)^2}{\sinh^2(x/2)} \; ; \quad x = \frac{\hbar\omega}{k_B T} \qquad (2b)$$

This function acts as a temperature-dependent low-pass filter with a cutoff frequency decreasing with decreasing temperature. Figure 2(b) shows that at low temperatures, $f_C$ again "selects" only low-frequency modes (at 150 K, $f_C$ drops to half its maximum at ~9.15 THz). When the temperature rises, e.g. to room temperature, also phonons at intermediate frequencies play a role.

The situation changes, when in addition to phonon occupation also other (potentially material dependent) factors play a role. In the following, this is exemplarily shown for the thermal conductivity tensor, κ, of naphthalene derived from the linear Boltzmann transport equation. Eq 3a shows, how κ can be calculated from the material's harmonic elastic properties, summarized in the tensorial quantity, $\eta^\lambda$, and from the phonon lifetimes, τ (as intrinsically anharmonic properties).

$$\kappa_{ij} = \frac{k_B}{N_q V} \sum_{\lambda=(n,q)} \eta_{ij}^\lambda \, \tau_\lambda \qquad (3a)$$

$$\eta_{ij}^\lambda \, k_B = f_C(\omega_\lambda, T) \left( v_{g,\lambda} \otimes v_{g,\lambda} \right)_{ij} \qquad (3b)$$

The summation index, λ, is a shortcut notation for band index *n* and wave vector *q*. As in the present contribution we are exclusively concerned with harmonic properties, in the following $\eta^\lambda$ shall be discussed in somewhat more detail. It is defined by eq 3b and contains the dyadic product of the group velocities $v_g$ as well as the heat capacity weighting function $f_C$ (see eq 2b). The *xx* component of $\eta^\lambda$ is plotted in Figure 2(c) for naphthalene (for the DFT reference calculation; see Section 3.2.1). Here one sees that, due to a pronounced decrease of the group velocities with frequency, the contributions to the thermal conductivity are more strongly confined to low frequencies than the contributions to the heat capacity.

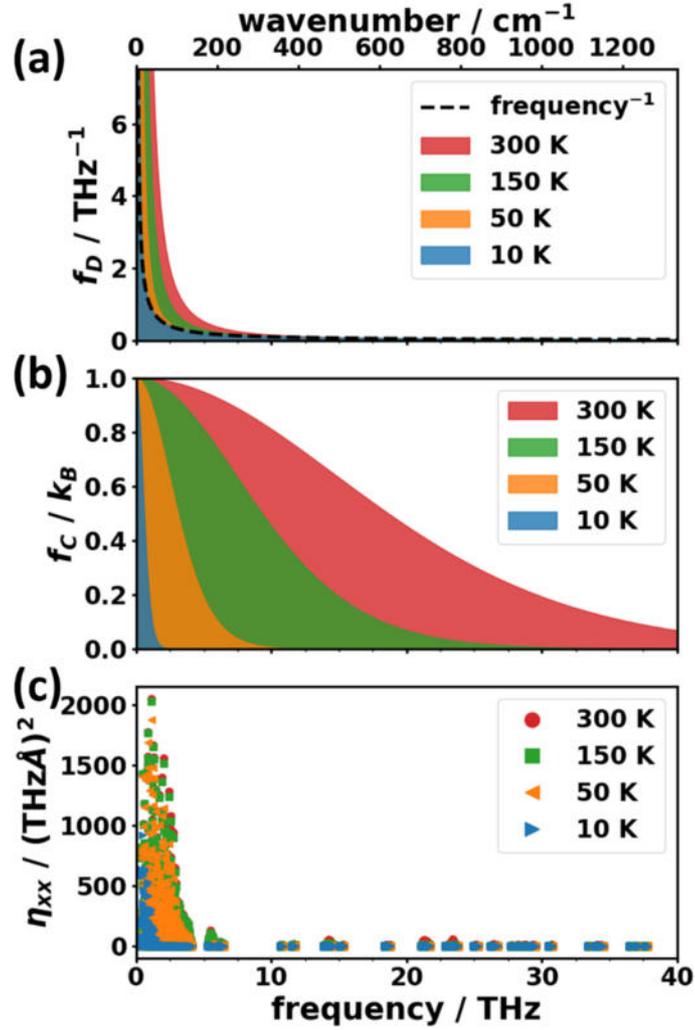

*Figure 2.* Frequency dependence of the material independent spectral function for (a) the mean squared thermal displacements $f_D$ and (b) the phonon mode heat capacity $f_C$. (c) contains the harmonic part of the mode contributions $\eta_{xx}$ to the thermal conductivity (xx component). For further details on the plotted quantities see main text.

There are, however, also quantities for which the contributions of all phonon modes are relevant. An example for such a quantity is the free-energy, which determines, e.g., the relative stability of polymorphs. Here, as a consequence of the zero-point energy, not only the contributions of occupied modes count. Notably, for the zero-point energy high-frequency modes are particularly relevant.

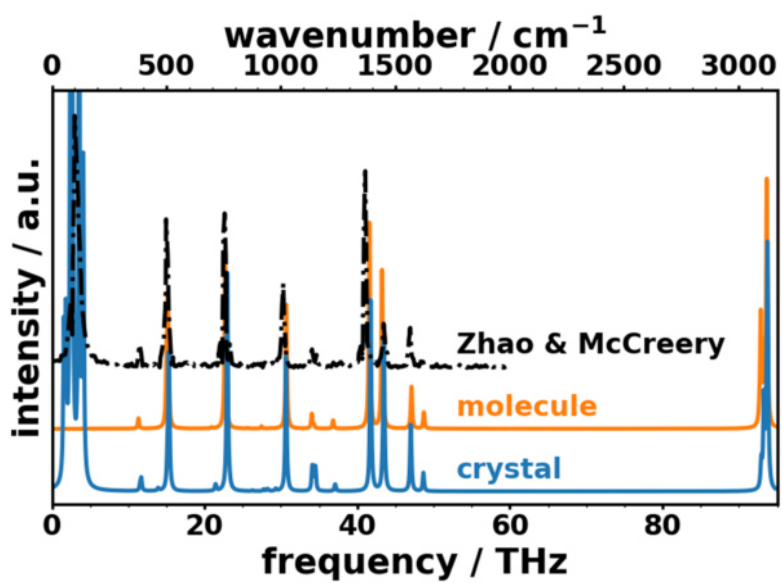

*Figure 3.* Simulated unpolarized Raman spectra for molecular (calculated with Gaussian16, 6-311G(d,p)++/PBE) and crystalline (VASP, PBE) naphthalene (solid lines) compared with experimental data from Zhao and McCreery[73]. For both simulated spectra and the measurement of Zhao and McCreery, an excitation wavelength of 784 nm was used. The nature of the morphology is not specifically described in ref 73, but was privately communicated to us by Richard McCreery.

An additional aspect that makes a discrimination between a low- and a high-frequency region advisable is that the nature of the vibrations in the two regimes are fundamentally different in organic semiconductors. This can be seen when comparing Raman spectra of naphthalene calculated for an isolated molecule with those of the molecular crystal: the calculated spectrum for the crystal display an excellent over-all agreement with the experimental data on polycrystalline naphthalene by Zhao and McCreery[73], (see Figure 3). Above ~10 THz the molecular and crystal spectra agree very well, suggesting that also in the crystalline environment the respective vibrations are primarily of intramolecular nature. Conversely, the pronounced features below ~5 THz appear only for the molecular crystal, which implies that they are associated with intermolecular motions.

An analysis of the eigenmodes indeed shows that the twelve lowest-lying bands (up to ~4.0 THz) correspond to motions in which the two naphthalene molecules in the unit cell essentially rotate or translate relative to each other.

## 3.2 Phonons in the low-frequency regime (up to 9 THz)

### 3.2.1 DFT simulations: Identifying a suitable reference methodology

In view of the discussion in the previous section, we will first analyze the performance of the various methodologies in the low-frequency regime. Since the details of the phonon band structure of naphthalene has already been discussed, e.g., in refs 25,30,74, in the following, we will primarily focus on the impact of different levels of theory on the obtained numerical results. As a first step, we will test to what extent DFT calculations relying on the PBE functional can serve as reference calculations for more approximate approaches. This would be useful, as the available experimental phonon band structures are not sufficient for generating reference data for thermodynamic quantities such as free energies and heat capacities. This is because the they were only measured in selected high-symmetry directions and up to 4 THz (i.e., only for the intermolecular modes). Bearing in mind that the low-frequency vibrations are crucially impacted by intermolecular interactions (see above), for identifying the ideal reference methodology it is useful to first assess the performance of different *a posteriori* vdW correction schemes. In this context, it has been shown recently for Γ-point vibrations (specifically for Raman spectra) that the D3-BJ correction yields highly accurate results for a variety of rather complex organic semiconductors and their polymorphs[23]. Essentially the same accuracy has been obtained in ref 23 with the many-body dispersion (MBD) method[75,76], albeit at sharply increased computational costs. Nevertheless, employing the MBD approach would be elegant, as it is based on the expression for the correlation energy in the random phase approximation and, thus, does not only account for two-body, but also for many-body dispersion interactions. Notably, also for naphthalene, D3-BJ and MBD van der Waals corrections yield nearly perfect agreement both in terms of lattice parameters (see Table 1) and for Γ-point frequencies (see Supporting Information). As far as phonon band structures are concerned, we, have, however, not been able to converge the MBD simulations, which we, tentatively, attribute to the very large number of atoms in the supercells needed to calculate phonon bands (containing 432 atoms). In passing we note that Brown-Altvater[30] found that, when fixing the unit cell size to the experimental value, also disregarding van der Waals interactions provides a good agreement between measured and calculated phonon bands, similar to what we have observed for the above-mentioned analysis of low-frequency Raman spectra[23]. This approach, however, does not allow a meaningful optimization of the naphthalene unit cell and, thus, will also not be pursued in the following.

Consequently, Figure 4(a) contains only a comparison between the phonon bands calculated employing the D3-BJ approach and the experimental results measured at 6 K for deuterated naphthalene[18]. The overall agreement between theory and experiment is excellent, with a root-mean-square deviation (RMSD) between the measured data points and the calculated ones amounting to ~0.13 THz (~4.3 cm$^{-1}$). This is true for the entire frequency range for which experimental data are available (i.e., up to 4 THz). In this context it should be noted that an analysis of the associated displacement patterns reveals that this frequency range also comprises all six bands dominated by intermolecular vibrations. In the following comparison we will, however, also include the largely intramolecular bands between ~5 THz and 6 THz as a first benchmark for the performance of the employed methodology for describing the relative energetics of inter- vs. intramolecular phonons.

The two other *a posteriori* vdW corrections that have been tested in ref 23 in terms of reproducing reliable Raman spectra are the Grimme D2 approach[77] and the Tkatchenko-Scheffler (TS) method[78]. In contrast to the D3-BJ method, the D2 vdW correction does neither include an attractive $r^{-8}$-term in addition to the $r^{-6}$ term, nor do the coefficients depend on the atoms' local coordination. Additionally, the Fermi-type damping function in the D2 method is less smooth than the Becke-Johnson damping in D3-BJ. The TS approach is similar to the D2 method in terms of the functional form of the vdW interaction. It differs, however, in the way the vdW coefficients are calculated: while D2 uses tabulated values, the TS method calculates them based on the atomic polarizabilities derived from the associated Hirschfeld partitioning of the system's charge density. The phonon band structures obtained with those two vdW corrections are compared to the low-temperature (6 K) experimental data[18] in Figure 4(b) and (c). The respective optimized unit-cell parameters are also contained in Table 1. Interestingly, the unit-cell parameters from the TS calculation still agree rather favorably with the experiments. This also applies to the acoustic phonon bands. For the optical bands, we, however, obtain significant deviations between the TS simulations and the experiments, which increase with phonon frequency. The D2 approach underestimates the unit cell volume by ~10% with lattice vectors underestimated by ~0.1 – 0.2 Å.

In spite of this apparent overestimation of the van der Waals attraction, for the phonon band structure the agreement with experiments [Figure 4(c)] is significantly better than in the TS case (but at the same time significantly worse than for the D3-BJ approach).

The methodology-dependent shifts of the bands are also reflected in the low-frequency DOSs [see Figure 4(d)], which represents the situation in the entire 1BZ rather than along the high-symmetry paths in the band structures. Comparing the DOSs, one can again see that the TS approach shifts many spectral features to higher frequencies. This also applies to the lower edge of the band gap occurring between the highest intermolecular and lowest intramolecular modes (at ~4 THz in the D3-BJ results). This results in a nearly complete closure of the gap in the TS calculations. In D2 this shift is less pronounced and the band gap is reproduced relatively well. However, the peaks in the D3-BJ-DOS below 2 THz are entirely washed out in the D2 calculation, suggesting that the bands are more "homogeneously" distributed than in D3-BJ.

*Table 1.* Comparison of simulated lattice constants of (non-deuterated) naphthalene with experimental data (measured at 5 K[a])[b].

|  | |a| / Å | |b| / Å | |c| / Å | $\beta$ / ° | unit cell volume / Å3 |
|---|---|---|---|---|---|
| Experiment[a] | 8.080(5) | 5.933(2) | 8.632(2) | 124.65(4) | 340.41 |
| DFT + D3-BJ | 8.078 | 5.903 | 8.622 | 124.24 | 339.91 |
| DFT + MBD | 8.090 | 5.910 | 8.608 | 124.24 | 340.25 |
| DFT + TS | 8.052 | 5.860 | 8.616 | 123.89 | 337.47 |
| DFT + D2 | 7.822 | 5.821 | 8.485 | 125.34 | 315.09 |
| DFTB | 7.573 | 5.733 | 8.457 | 125.04 | 300.61 |
| COMPASS | 8.002 | 5.771 | 8.500 | 124.64 | 322.96 |
| MOF-FF | 7.998 | 5.884 | 8.635 | 123.18 | 340.18 |
| GAFF | 7.850 | 5.979 | 8.610 | 124.05 | 334.88 |

[a]The experimental data were taken from ref 74. [b]The digits in brackets in the experimental data imply the measurement uncertainty.

These comparisons show that DFT/D3-BJ is clearly the best suited "high-level" theoretical methodology that can be applied to benchmark the other more approximate approaches considered in this work. Thus, DFT/D3-BJ results will be referred to as "DFT ref" in the following.

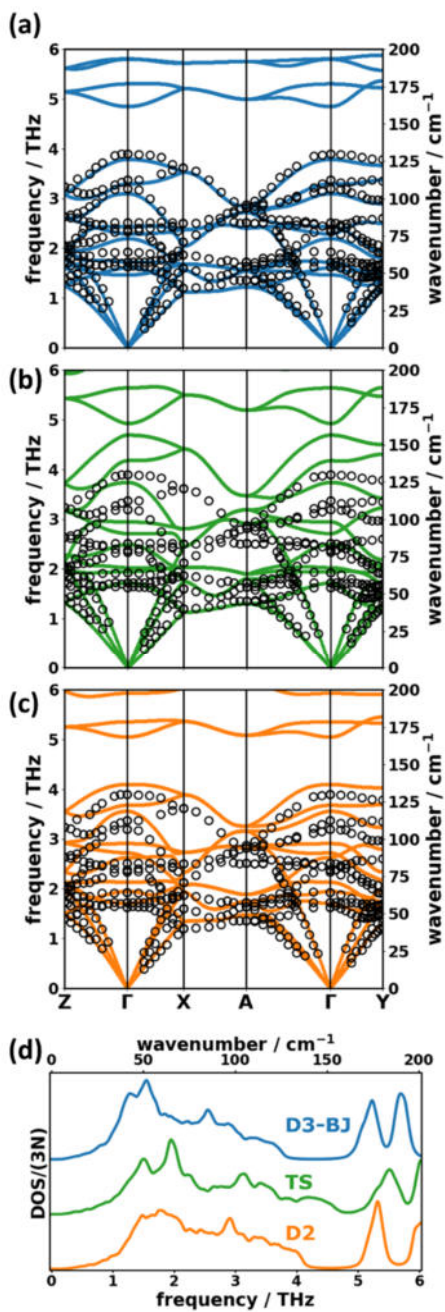

*Figure 4.* *Phonon band structure of deuterated naphthalene simulated with the (a) D3-BJ, (b) TS, and (c) D2 vdW correction compared to data measured by inelastic neutron scattering at 6K (open*

*circles)[18]. Panel (d) compares the normalized DOSs for the three theoretical methodologies. The frequency range contained in the plots does not cover the entire low-frequency region defined earlier. This is because there are no experimental data above 4 THz and the deuteration of naphthalene shifts all calculated frequencies to lower values compared to the non-deuterated molecules. We also note that a comparison of the experimental phonon dispersion of deuterated naphthalene with simulations based on DFT employing different functionals can be found in ref 30.*

Up to this point, we have considered deuterated naphthalene in order to validate the DFT reference data comparing them to the neutron scattering data. In the following it is no longer necessary to maintain the deuteration. Thus, from now on all displayed data describe the practically more relevant non-deuterated species.

### 3.2.2 Obtaining phonon band structures with density functional tight-binding theory

A significant speedup of the computations can be achieved by switching to a more approximate methodology, like density functional tight-binding (DFTB). This, however, comes at a cost: as can be seen in Table 1, the DFTB calculated lattice parameters differ quite significantly from the experiments and from the DFT reference calculations. Notably, |*b*| and |*c*| are slightly shorter by approximately the same amount (~2-3%), but |*a*| is shorter by ~6%. This results in a much denser herringbone packing. As the inter-molecular vibrations are particularly sensitive to the packing structure, one can expect deviations for the low frequency modes. These are indeed observed in the calculated band structure and the DOS in Figure 5(a): the band widths and slopes of the acoustic bands in DFTB are significantly overestimated compared to the DFT reference, which results in a more gradual onset of the associated DOS together with a shift of its peaks to higher energies. This can be understood as a direct consequence of the denser packing, resulting in an increased mechanical stiffness of the system. Also for the optical bands massive differences are observed. Some of the bands are primarily shifted to higher frequencies, which results in a shift of the DOS peaks. In many cases also the band shapes change dramatically. Most significantly, the band gap between the highest intermolecular and the lowest intramolecular modes observed in the DFT calculations between ~4.1-5.3 THz disappears for DFTB (similar to the above-described DFT/TS case). The closing of the gap in DFTB is mostly a consequence of an upwards shift of the two intermolecular bands at the lower edge of the gap (~ 3.56 THz and 4.11 THz in DFT, and ~5.14 THZ and 5.56 THz in DFTB at Γ). An analysis of the associated displacement patterns shows that the vibrations correspond to molecules essentially twisting around their long axes either out of phase (lower band) or in phase (higher band). Not surprisingly, the increased packing density for the DFTB geometry affects these modes particularly strongly.

This raises the question, whether the deviations between the DFT and the DFTB bands could be fixed by using more realistic unit cell parameters. In fact, Brandenburg and Grimme[48] suggested that for calculations of phonon bands employing DFTB, one should still rely on the lattice parameters obtained in a DFT optimization and only optimize the atomic positions at the DFTB level. The low-frequency bands obtained with this approach (referred to as "DFTB@DFT") are compared to the DFT reference data in Figure 5(b).

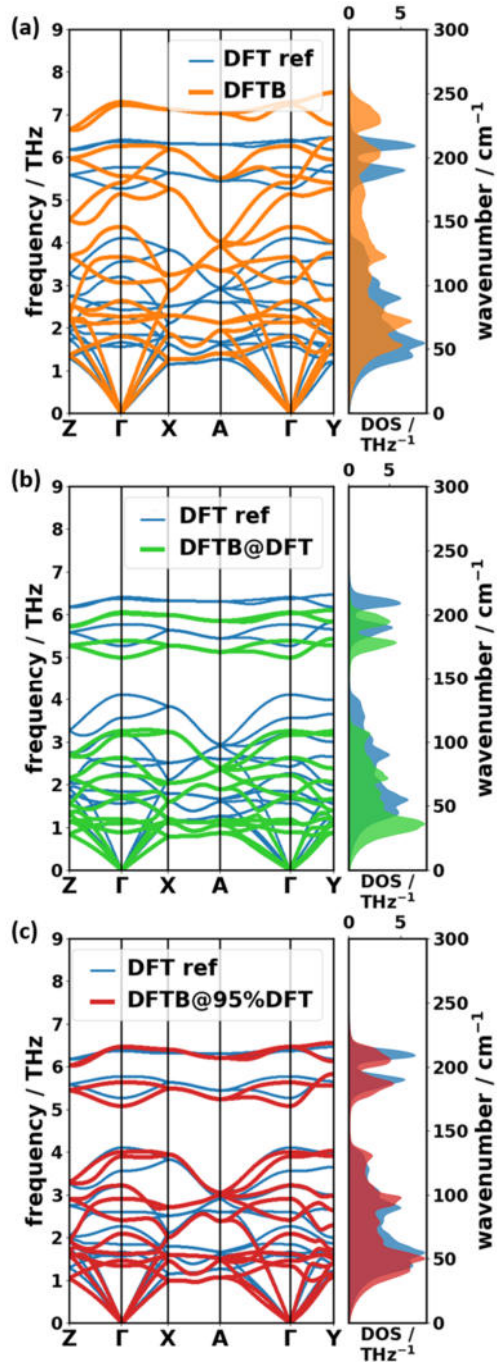

*Figure 5*: *Comparison of phonon band structures and DOSs of the reference DFT/D3-BJ calculation (DFT ref) with DFTB/D3-BJ simulations based on (a) a structure with DFTB-optimized unit cell, (b) a structure with the DFT/D3-BJ unit cell, and (c) a structure calculated for the DFT/D3-BJ unit cell isotropically rescaled by a factor of 0.95.*

Indeed, certain aspects of the phonon bands are improved. For example, the first band gap and the shape of the DFTB@DFT DOS is more similar to the DFT reference. The quantitative agreement is, however, still far from satisfactory. Compared to the DFT reference, the energy scale is now compressed rather than expanded; i.e., phonons are shifted to too low frequencies. Like for most materials, the mode Grüneisen parameters in naphthalene have a positive sign[25,79,80] (i.e., phonon frequencies increase upon decreasing the unit cell volume). This suggests that a solution to too low frequencies would be decreasing the volume of the primitive unit cell. Indeed, the best result for a DFTB calculated phonon DOS and band structure is found when rescaling the unit-cell volume to 95 % of the DFT volume ["DFTB@95%DFT", Figure 5(c)]. This rescaling minimizes the RMS deviation (RMSD) of frequencies with respect to the DFT reference regardless of whether modes from the entire 1BZ or only at Γ are considered (see Supporting Information). This suggests that a possible strategy for obtaining quantitatively more accurate DFTB bands could be to do determine the optimum unit cell rescaling factor based on Γ-point frequencies (where DFT/D3-BJ calculations are affordable also for more complex molecular crystals). This "optimized" unit cell could then potentially be used as the basis for DFTB calculations of the phonon bands, but before generally applying that approach, tests on alternative systems would be advisable. Moreover, at finite temperature one would have to find the scaling factor that matches the phonons at the thermally expanded volume as, anharmonicities are not necessarily equally described in DFTB and DFT. Thus, a general application of this procedure is far from straightforward.

### 3.2.3  Obtaining phonon band structures with classical force fields

The largest speed-up of the calculation of phonon bands can be achieved by describing interatomic interactions with parametrized force fields (FFs). Interestingly, independent of their level of sophistication all used force fields (COMPASS, MOF-FF, and GAFF) yield optimized unit cells, whose lattice parameters are in close agreement with the DFT reference data and the experiments (see Table 1). In particular for MOF-FF, the unit cell volume is essentially identical to the references in spite of the fact that the force field parametrization has been performed on an isolated molecule. Moreover, dispersion interactions, which are particularly relevant for inter-molecular interactions, are not part of the parametrization process at all. Using MOF-FF and GAFF, a maximum relative deviation of ~-2% and ~-3% is observed for $|a|$, which determines the distance between the two inequivalent molecules in the unit cell. In contrast, in the COMPASS optimizations the largest deviations (~-2%) are found for the $b$ lattice parameter – i.e. the short distance between one molecule and its periodic image. The use of force fields in the geometry optimization also results in a less symmetric geometry (COMPASS and GAFF: space group P1, MOF-FF: space group P$\bar{1}$) compared to the one obtained in the reference calculation and in the experiments (space group P2$_1$/a).

The comparison of the band structures and DOSs of the FFs with the DFT reference data is shown Figure 6. The strength of the COMPASS force field lies in an accurate description of the acoustic bands. The DOS up to ~1.2 THz shows the best agreement with the reference for all cases discussed so far, which results in an onset of the DOS perfectly matching the DFT reference. At higher frequencies, however, the agreement deteriorates. The edges of the band gap are relatively far from the reference, and the higher bands (~ 4.0 – 6.5 THz) are not only shifted, but also show significant changes in their dispersion. This particularly applies to the intramolecular bands, where the parametrization of the bonding interactions of the force field is expected to play a crucial role.

The overall agreement with the reference data is clearly improved for our MOF-FF parametrization of naphthalene [see Figure 6(b)], with a satisfactory reproduction of most of the characteristic features of the reference DOS and band structure. Only the bands below ~2 THz are slightly shifted to lower frequencies resulting in a premature rise of the phonon DOS. The better performance of MOF-FF compared to COMPASS for the intramolecular bands is not unexpected, considering that especially the bonding part of MOF-FF (together with the atomic charges) has been parametrized/fitted based on a DFT calculation of a naphthalene molecule, while the COMPASS potential is not system-specific.

Interestingly, GAFF provides an excellent agreement with the reference for the low-frequency band structure and DOS in spite of its very simple structure. Most of the characteristic features in the band dispersion are reproduced, although some bands experience shifts to (typically higher) frequencies. In fact, up to 4.5 THz the agreement with the reference is nearly as good as for the MOF-FF calculations, and for the acoustic bands, GAFF even outperforms MOF-FF. This supports the notion that the non-bonding interactions (vdW and Coulomb) are most relevant for the intermolecular modes in this spectral range and that their description in GAFF occurs at an acceptable level of accuracy. As far as the intramolecular bands above ~5 THz are concerned, the agreement between GAFF and the DFT reference clearly deteriorates, which can be attributed to the purely harmonic bonding interactions in GAFF and the omission of any cross terms.

Concluding this section on the simulation of band structures, it should be remarked that the same trends described above for the comparison relative to the DFT reference are also found, when comparing the bands to the experiments. This is explicitly shown in the Supporting Information.

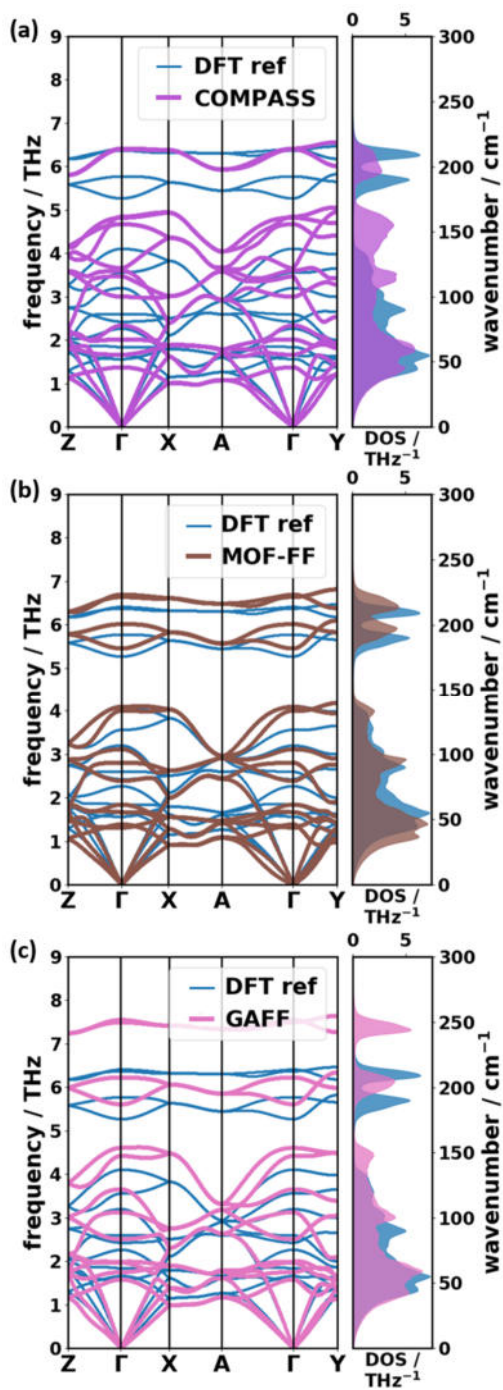

***Figure 6.*** *Comparison of phonon band structures and DOSs of the reference calculation (DFT ref) with the results obtained using the following force fields for both, the unit cell optimization and the frequency calculation: (a) COMPASS[54], (b) our own paramtrization of the MOF-FF [61], and (c) GAFF[49].*

### 3.2.4 Quantitative benchmark of phonon properties in the low-frequency region

In order to obtain more quantitative insights into the performance of the different methods, it is useful to analyze bands not only along high-symmetry directions, but rather to sample the entire 1 BZ. As described in Section 2.4, this is done by employing a 9×10×9 $q$-mesh. Figure 7(a) shows the deviations of the calculated phonon frequencies between the more approximate methods and the DFT/D3-BJ reference data obtained in this way. In passing we note that to make such a comparison it is necessary to identify equivalent phonon modes calculated with different approaches. For that, we relied on the respective phonon eigenvectors (polarization vectors) and made use of the algorithm of Kuhn[81] as described in more details in the Supporting Information.

To obtain quantitative descriptors for the performance of a specific method for calculating a quantity $x$, we calculated the root mean square deviations (RMSD$_x$), as defined in eq 4a.

$$RMSD_x = \sqrt{\frac{1}{N}\sum_i^N (x_i - x_{ref,i})^2} \qquad (4a)$$

Unfortunately, the RMSD$_x$ values do not provide information, whether a method typically under- or overestimates a given quantity. Moreover, large values of the RMSDs for quantities associated with the phonon band structure (like frequencies or group velocities) do not necessarily mean that derived quantities (like heat capacities) are poorly described. For those, one might encounter fortuitous error compensations between frequency ranges in which, e.g., frequencies are overestimated and regions in which they are underestimated. To assess, whether a method is prone to that, we will compare RMS deviations (RMSD) to average deviations AD$_x$ defined in eq 4b.

$$AD_x = \frac{1}{N}\sum_i^N x_i - x_{ref,i} \qquad (4b)$$

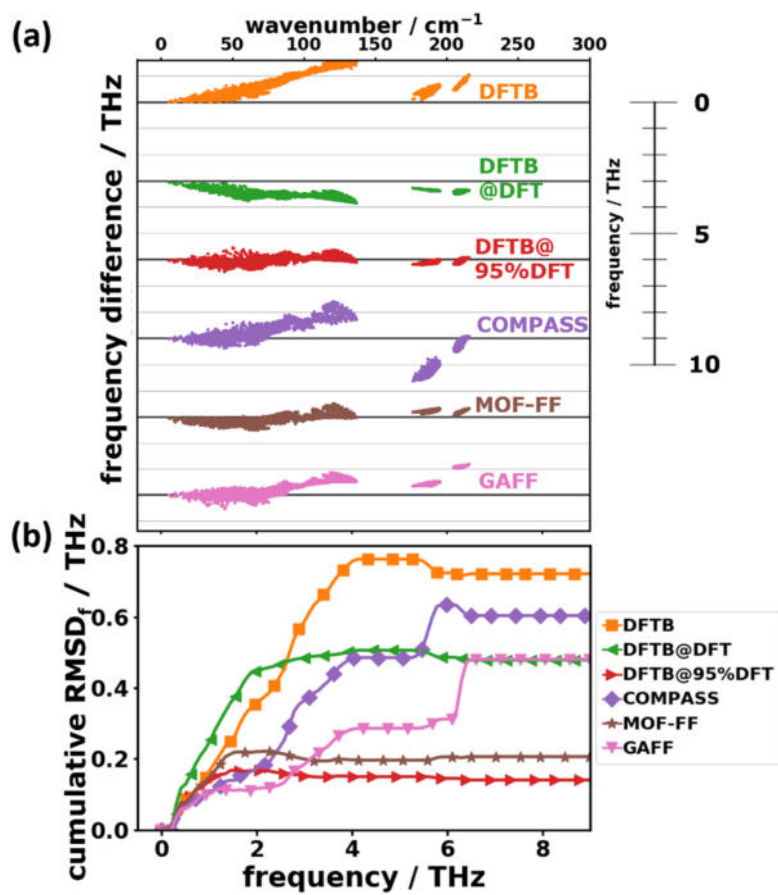

*Figure 7.* (a) Frequency differences with respect to the reference (DFT/D3-BJ) for the various approaches as a function of the reference frequency in the low-frequency region. Each approach has its own zero line (thick black horizontal lines). (b) Cumulative root mean square deviations of frequencies below a certain cutoff frequency as a function of that cutoff frequency.

As far as the phonon frequencies are concerned, Figure 7(a) shows the calculated deviations relative to the DFT/D3-BJ reference data (DFT ref). Figure 7(b) contains the evolution of the cumulative $RMSD_f$, i.e., the $RMSD_f$ as a function of the frequency up to which the summation in eq 4a has been performed. The "final" values for the entire low-frequency range (i.e., up to 9 THz) are shown in Table **2** together with the corresponding average deviations $AD_f$. Consistent with the conclusions drawn from the band structures, in DFTB most modes experience shifts to higher energies by up to ~1.5 THz, with the error increasing roughly linearly with frequency up to ~4 THz. This causes a sharp rise of the associated $RMSD_f$ in that frequency region and a rather large positive value of the $AD_f$. Conversely, the DFTB@DFT frequencies are generally too small (by up to ~0.9 THz) resulting in a negative $AD_f$. Below 2 THz, the corresponding cumulative $RMSD_f$ is even larger than for DFTB. The frequency differences of associated modes are significantly decreased by rescaling the unit cell (DFTB@95%DFT).

Using the COMPASS FF, one tends to overestimate the intermolecular frequencies up to 4 THz and underestimates the frequencies of the two intramolecular bands around ~6 THz. The majority of the modes can be found within ~±1.5 THz around the reference. The $RMSD_f$ value continuously increases for the intermolecular modes and then experiences a step for the intramolecular vibrations due to their particularly bad description. The comparably small value of the $AD_f$ for the COMPASS force field is a clear indication for an "error compensation" between inter- and intramolecular modes (see below). With MOF-FF, the frequency spread can be reduced significantly to a level comparable to DFTB@95%DFT. This is also manifested in the evolution of the $RMSD_f$ Figure 7(b). Consistent with a particularly small value of the $AD_f$, there is no systematic over- or underestimation of frequencies. In line with the observations for the band structures, phonon frequencies from the GAFF calculations perfectly fit the reference data up to ~2.5 THz. This is also visible in Figure 7(b), where one sees that as far as the value of the $RMSD_f$ is concerned, GAFF outperforms MOF-FF in a spectral region of up to ~3 THz. However, at higher frequencies, the evolution of the cumulative $RMSD_f$ suffers from an overestimation of phonon energies (between ~3.0 THz and 4 THz, and around 6 THz).

As a next step, we will analyze the phonon group velocities. Here, first insights can be gained from the density of group velocities (DOVG), i.e. the number of phonons with a given group velocities per unit $v_g$ interval. The DOVGs are shown in Figure 8. Although the overall trend in the DOVG is similar for all approximate methods and the DFT reference, the relative deviations are quite sizable especially for large group velocities. Their densities are massively overestimated by all methods apart from DFTB@DFT. Concomitantly, the densities at small group velocities are typically underestimated. A disadvantage of such a "density of group velocities" analysis is that it does not take into account for which mode a specific group velocity is calculated.

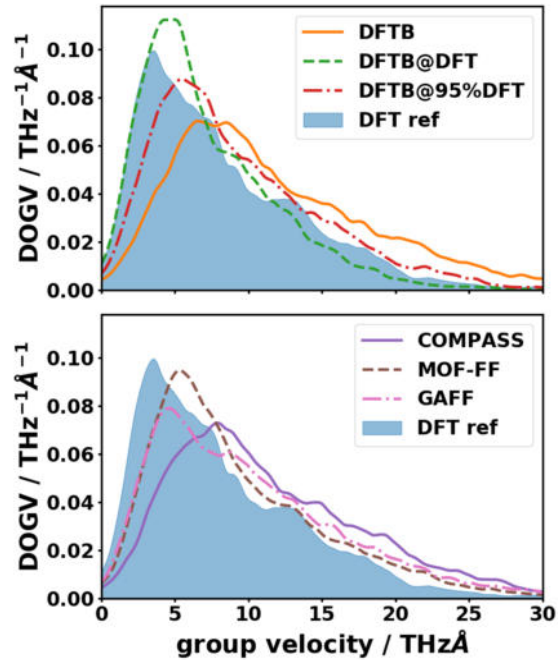

*Figure 8. Density of group velocities (DOGV) as a function of the norms of the group velocity vectors. The DOGV has been calculated as a sum of Lorentzian peaks (FWHM = 1 THzÅ) centered at the group velocity of each low-frequency (≤ 9 THz) phonon mode on a discrete mesh normalized by the total number of considered modes. The shaded area represents the DFT reference.*

Thus, we applied a similar statistical analysis as for the frequencies to the norms of the group velocities. This allows to quantitatively assess the agreement in the band dispersions. The cumulative $RMSD_{v_g}$ in the low-frequency regime is shown in Figure 9 together with the associated mode group velocities.

Here, the best agreement with the reference data for the entire low-frequency region is obtained for the DFTB@DFT calculations (lowest value of the cumulative $RMSD_{v_g}$ at 9 THz - see also Table **2**– consistent with the best matching DOGV in Figure 8). However, for this method the cumulative $RMSD_{v_g}$ is particularly high in the region of the acoustic bands up to ~2 THz and this is the region most relevant when studying thermal transport. In that frequency region, MOF-FF displays the smallest deviations.

Overall, with the exception of the DFTB and DFTB@DFT approaches, the values of the $RMSD_{v_g}$ do not change significantly between 1 THz and 9 THz indicating that there is no significant variation in the performance for the different acoustic optical bands. Moreover, with the exception of DFTB@DFT all methods have the tendency to overestimate the group velocities, as already inferred from the DOGVs. A comparison (in Figure 9) of the $RMSD_{v_g}$ values and the absolute values of the group velocities reveals that both quantities are of the same order of magnitude. Consequently, as far as group velocities are concerned, none of the approximate methodologies displays a fully satisfactory performance.

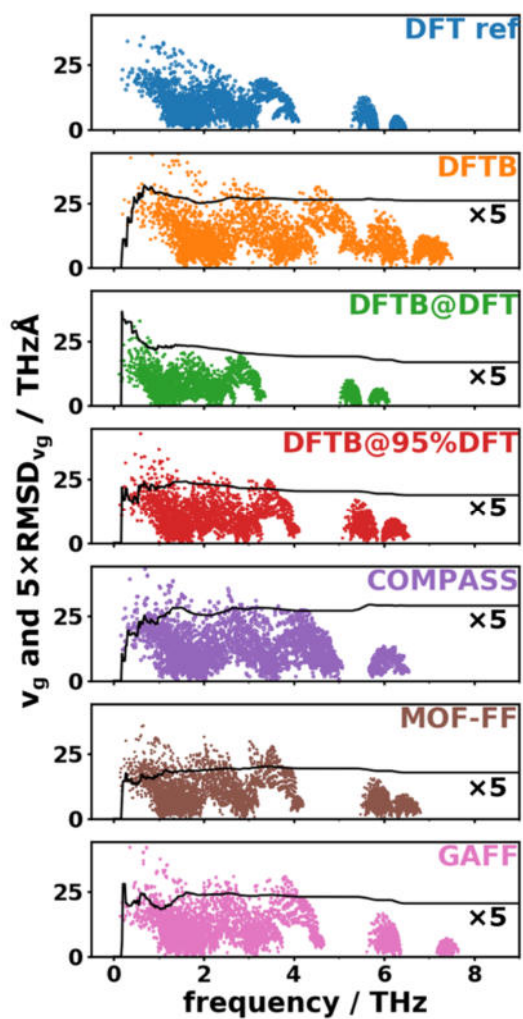

***Figure 9.*** *Group velocity of each phonon modes on the discrete mesh described in section 2.4 as a function of its frequency. The solid black lines show the cumulative $RMSD_{v_g}$ as a function of frequency. Note that, for the sake of comparison, the values of the $RMSD_{v_g}$ have been multiplied by a factor of 5.*

**Table 2.** Mean differences in various quantities $AD_x$ ($x$…frequencies $f$, norm of the group velocity vectors $v_g$, and mean squared thermal displacement, MSTD) and RMS deviations (RMSD) with respect to the DFT reference[a].

| | | DFTB | DFTB@DFT | DFTB@95% DFT | COMPASS | MOF-FF | GAFF |
|---|---|---|---|---|---|---|---|
| Frequencies ($\leq$ 9 THz) | $RMSD_f$ / THz | 0.72 | 0.48 | 0.14 | 0.60 | 0.20 | 0.48 |
| | $AD_f$ / THz | 0.62 | -0.46 | -0.05 | 0.07 | -0.03 | 0.31 |
| Frequencies (entire range) | $RMSD_f$ / THz | 1.43 | 1.50 | 1.46 | 2.10 | 0.76 | 4.14 |
| | $AD_f$ / THz | -0.55 | -0.84 | -0.73 | 0.17 | 0.06 | 1.10 |
| Group velocities ($\leq$ 9 THz) | $RMSD_{v_g}$ / THzÅ | 5.2 | 3.3 | 3.8 | 5.8 | 3.6 | 4.1 |
| | $AD_{v_g}$ / THzÅ | 3.8 | -0.9 | 1.2 | 3.3 | 1.6 | 1.7 |
| MSTD (150 K) | $RMSD_{u^2}$ / Å² | 0.024 | 0.076 | 0.019 | 0.005 | 0.033 | 0.002 |
| | $AD_{u^2}$ / Å² | -0.023 | 0.074 | -0.019 | -0.002 | 0.033 | 0.001 |
| MSTD (300 K) | $RMSD_{u^2}$ / Å² | 0.048 | 0.152 | 0.039 | 0.009 | 0.066 | 0.004 |
| | $AD_{u^2}$ / Å² | -0.046 | 0.147 | 0.037 | -0.004 | 0.066 | 0.003 |

[a]The listed values for frequencies and group velocities have been calculated from phonon modes sampled on a $q$-mesh in the 1BZ. The parameters for MSTDs are compared atom-wise to the reference.

### 3.2.5 Analyzing the performance of approximate methods for describing observables derived from the phonon band structures

As indicated in Figure 2 in Section 3.1, several physical observables nearly exclusively depend on the low-frequency modes. Therefore, in the following we will analyze how these observables are impacted by the deviations in the phonon band structures between approximate methods and the DFT/D3-BJ reference data. The first of these observables is the mean squared thermal displacement (MSTD) of the atoms (see eq 1a in Section 3.1). It gives a measure of the average variation in the atomic positions due to the thermal motion of the atoms. Figure 10(a) shows that the values of $\langle ||u_j||^2 \rangle$ somewhat vary for the chemically inequivalent carbon atoms. The same occurs for inequivalent hydrogens, for which due to their reduced mass on average the $\langle ||u_j||^2 \rangle$ are larger by ~70 %.

As far as the performance of the approximate methods is concerned, Figure 10(a) shows that by far the best agreement with the DFT/D3-BJ reference is obtained for the COMPASS and GAFF force fields. This is confirmed by the particularly small values for the RMSD$_{u2}$ listed in Table **2**. The DFTB values underestimate the mean squared thermal displacements (resulting in a negative value of the AD$_{u2}$), while DFTB@95%DFT, MOF-FF, and DFTB@DFT increasingly overestimate them (in line with an increasingly positive value of the AD$_{u2}$). This overestimation can become rather large such, that the $\langle ||u_j||^2 \rangle$ values for DFTB@DFT become roughly twice as large as the DFT/D3-BJ values. To understand that, one has to realize that the quality of the description of the $\langle ||u_j||^2 \rangle$ directly correlates with the quality of the description of the acoustic phonons: One reason for that is that at a given temperature these phonons display the highest occupation numbers n(ω,T). Additionally, the contributions of individual phonon modes to the $\langle ||u_j||^2 \rangle$ are scaled with a factor of 1/ω (c.f. eq 1a, 1b in Section 3.1). That scaling can be rationalized based on a comparison between a classic and a quantum-mechanical harmonic oscillator, as detailed, for example, in ref 82.

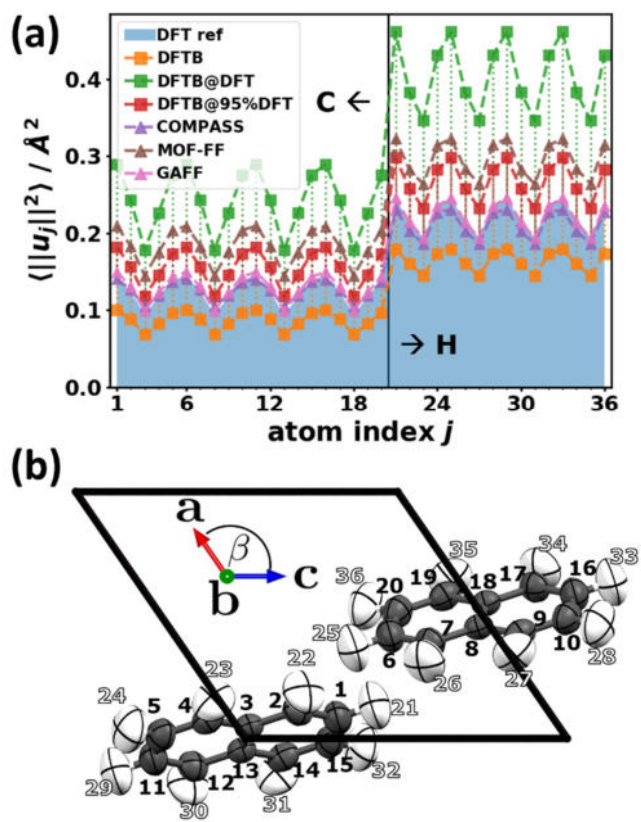

***Figure 10.*** *(a) Atom-resolved mean squared thermal displacements (MSTDs) $\langle||u_j||^2\rangle$ at 300 K. The shaded area represents the DFT reference. Vertical dashed lines and the lines connecting the MSTDs are guides to the eye to emphasize the difference with respect to the reference. (b) Atom labelling scheme in the naphthalene unit cell. In order to be able to graphically represent the thermal atomic motion, in crystallography one uses spatial Gaussian probability distributions with the MSTD being related to the covariance matrices. One then connects the points in space that lie on a chosen probability level to draw the thermal ellipsoids[83]. Here, the thermal ellipsoids are plotted with Mercury[84]) for the 75 % probability level (i.e., the probability for finding the atoms within the ellipsoids is 75 %).*

Thus, the excellent performance of the COMPASS and GAFF force fields (Figure 10) can be traced back to the perfect match of the corresponding onsets of the DOSs in the region of the acoustic phonons with the reference data (see Figure 6). The significant underestimation of the DOS in the acoustic region for DFTB (due to an overestimation of the frequencies of the acoustic phonons; see Figure 5), results in a significant drop in the equilibrium occupation of the acoustic phonons. This causes the significant decrease of $\langle ||u_j||^2 \rangle$. Conversely, for DFTB@95%DFT, MOF-FF, and DFTB@DFT, the frequencies of the acoustic phonons are underestimated such that their equilibrium occupation becomes much too high, causing the overestimation of the thermal displacements. As far as the temperature dependence of the deviations in the $\langle ||u_j||^2 \rangle$ is concerned, one sees that they approximately double upon increasing the temperature from 150 to 300 K (

Table **2**); i.e., while the relative deviations stay approximately the same.

A further relevant thermodynamic quantity that can be directly calculated from the phonon bands, is the phonon heat capacity $C_V$. As described in eq 2a in Section 3.1, $C_V$ can be calculated by integrating the DOS multiplied by a temperature-dependent low pass-like envelope function $f_C$. Thus, for very low temperatures only the low-frequency part of the DOS is considered in the integration of the heat capacity. In fact, as discussed in Section 3.1, at a temperature of ~150 K, the envelope function reaches half the value of its maximum for a frequency of 9 THz. Thus, Figure 11(a), shows the temperature dependence of the heat capacity (per unit cell) up to that temperature, normalized by its value given by the Dulong Petit law in the classical limit ($3Nk_B$, with $3N$ being the number of degrees of freedom per unit cell). To more easily judge the performance of the different approaches, Figure 11(b) explicitly shows the deviations from the DFT-calculated reference values.

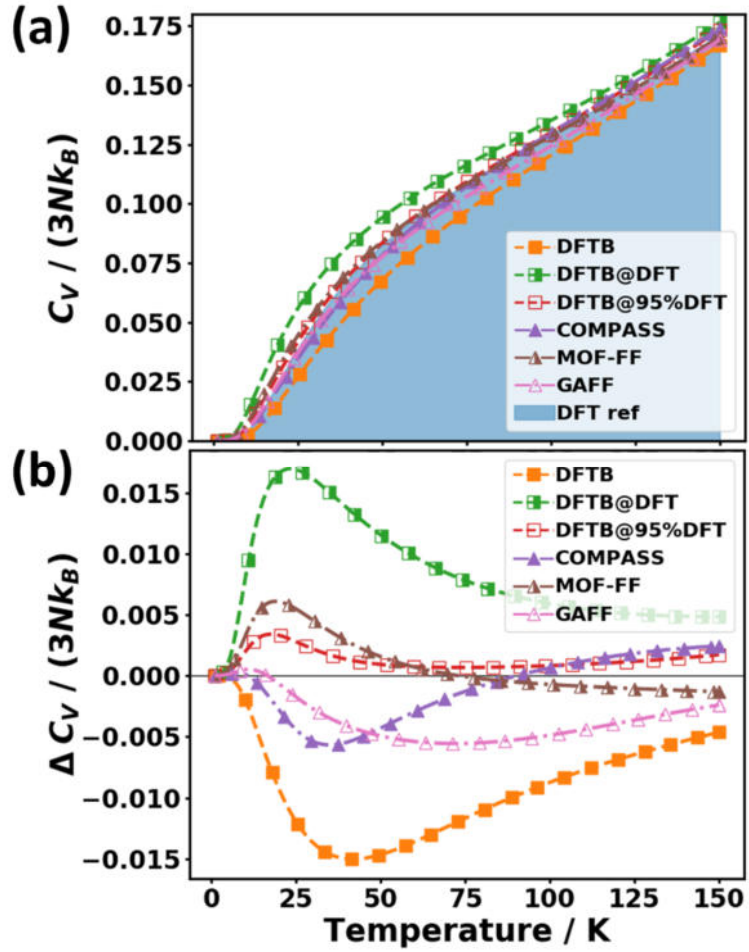

*Figure 11. (a) Phonon heat capacity $C_V$ normalized by the number of degrees of freedom 3N and the Boltzmann constant $k_B$ as a function of temperature for the tested approaches. The shaded area represents the DFT reference. (b) Difference in heat capacity with respect to the reference data (DFT ref). The symbols do not represent actually calculated data points (these lie much more densely), but rather serve as guides to the eye.*

In all approaches, the deviations are largest for temperatures below ~75 K. Consistent with the trends already observed for the $\langle ||u_j||^2 \rangle$, DFTB underestimates the heat capacity in the entire displayed temperature range. Similarly, GAFF and COMPASS, which displayed the best

performance for the $\langle ||\boldsymbol{u_j}||^2 \rangle$, show the smallest deviations in the present case, but only up to temperatures around 15 K. The reason, why for $C_V$ (in contrast to $\langle ||\boldsymbol{u_j}||^2 \rangle$) the performance of these force fields deteriorates already at rather small temperatures lies in the different shapes of the "low-pass filters" $f_C$ (determining the heat capacity; c.f. eq 2b) and $f_D$ (determining the mean square thermal displacements; c.f. eq 1b). As shown in Figure 2, for a given temperature, $f_C$ extends to much higher frequencies than $f_D$. This can be rationalized by higher-energy phonons carrying more energy, and, thus, also contributing more strongly to the heat capacity provided that the corresponding states are occupied. At even higher temperatures the trend in the error of $C_V$ is reversed especially in the COMPASS case due to the severe underestimation of the frequencies of the lowest lying intramolecular phonons (see Figure 7). This results in a partial error compensation, which is also apparent from a very small value of the AD$_f$ for COMPASS (see Table **2**). Consequently, that force field displays a rather good apparent performance in the entire considered temperature range.

For DFTB@95%DFT and MOF-FF the comparably small differences in $C_V$ relative to the reference can be traced back to the generally rather accurate description of the phonon frequencies in the entire low-frequency range, again with some error cancellations between overestimated and underestimated phonon frequencies. The strongest deviations from the reference are observed for DFTB and DFTB@DFT, where DFTB seriously underestimates $C_V$, while DFTB@DFT overestimates it. This is consistent with the above-discussed rather severe overestimation (underestimation) of the phonon frequencies by DFTB (DFTB@DFT). Overall, the maximum relative errors of $C_V$ reach up to 20% at ~45 K in DFTB and even up to 43% at ~25 K in DFTB@DFT. The absolute errors are, however, still rather small (~1.6% of the saturation value) because at 150 K the heat capacity has reached only ~17% of its saturation value. Note that the latter is (hypothetically) approached only above ~3500 K (i.e., far above the melting temperature of naphthalene at 353 K[85]), This can be explained by the high-frequency C-H stretching vibrations (above 90 THz) requiring high temperatures to be covered by the envelope function $f_C$ (see Supporting Information).

**3.3 Phonons in the high-frequency regime (vibrations above 9 THz)**

The above discussion for the heat capacity already suggests that for certain thermodynamic quantities also higher frequency vibrations are highly relevant. This, for example, applies to heat capacities at elevated temperatures or to the relative stability of different phases (i.e., the corresponding free energies). Also the nature of the modes at higher frequencies changes fundamentally, as the modes above ~4.1 THz (in the DFT/D3-BJ reference) are mostly intramolecular in nature. Thus, for their description an accurate modelling of bonding interatomic interactions is crucial, while for the lowest-frequency vibrations primarily non-bonding intermolecular interactions counted (see above). Thus, the trends discussed in the following as well as the relative performance of the different methods will not necessarily correlate with the observations discussed in Section 3.2.

**3.3.1 Comparison to experiment**

Unfortunately, no experimental data on the phonon band structures are available for frequencies above 4 THz. Therefore, for benchmarking the theoretical methodology in the high-frequency region, we have to resort to a comparison with results from vibrational spectroscopy (i.e., restrict the comparison to Γ-point frequencies only). Figure 3 compares the simulated and measured Raman spectra of naphthalene for frequencies up to 60 THz. The excellent agreement suggests that the DFT/D3-BJ calculations can serve as a viable reference for the entire frequency region. Notably, the very good agreement is not only observed for the simulations of the crystal, but also for the isolated molecules provided that the same functional is employed. This is another indication for the mostly intramolecular character of the vibrations at higher frequencies. Consequently, the impact of the type of applied van der Waals correction is only minor (see Supporting Information). In passing we note that in contrast to that the used functional plays a more significant role, where tests employing the hybrid functional B3LYP[86,87] (instead of PBE) yield an overestimation of the frequencies and, thus, a worse agreement with experiments (see Supporting Information).

### 3.3.2 Comparison of phonon properties at higher frequencies

A plot comparing the DOSs calculated with all applied methods is shown in Figure 12 for the entire frequency range. In that figure several observations can be made: (i) the DFTB-based approaches yield rather similar phonon DOSs above 9 THz independent of the choice of the unit-cell size. This is again consistent with the intramolecular nature of the higher-frequency vibrations. (ii) The lower edge of the large band gap (~49-92 THz) in the DFTB approaches agrees very well with the DFT/D3-BJ reference. Conversely, COMPASS overestimates the frequencies of these vibrations with the effect being even intensified in GAFF. In contrast, MOF-FF somewhat underestimates the corresponding frequencies. (iii) The C-H stretching modes (high frequency modes above 90 THz) are described rather poorly in most approaches: DFTB and GAFF significantly underestimate those frequencies. This error decreases for COMPASS. MOF-FF is the only approximate approach that yields a satisfactory agreement with the reference for these modes, albeit with a somewhat increased frequency-splitting of the symmetry-inequivalent vibrations.

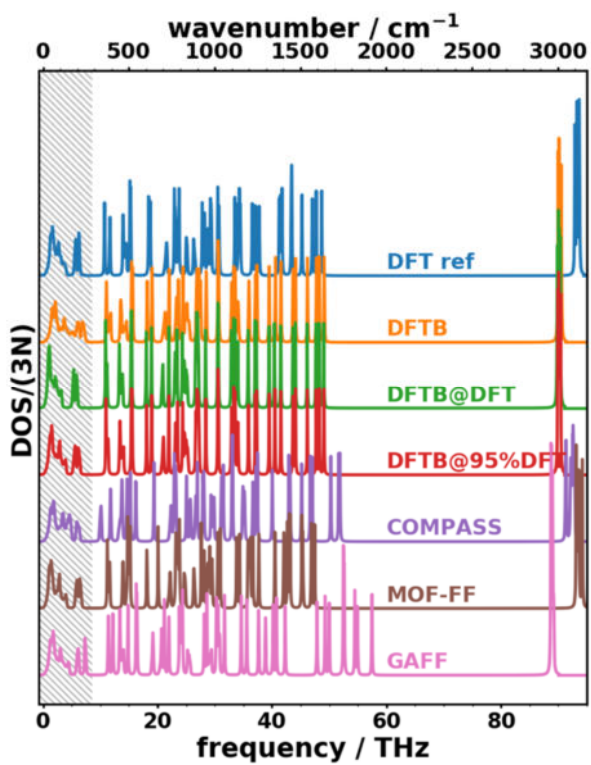

*Figure 12.* Densities of states as a function of phonon frequency of naphthalene for the various approaches tested in the current study. The hatched area highlights the low-frequency regime discussed in Section 3.2.

A shortcoming of comparing DOSs is that it provides information only on where in frequency vibrations exist but misses to assess the actual nature of these vibrations. To account for that, we again use the (complex) dot product between phonon eigenvectors of different simulation approaches to identify the most similar pairs of phonon modes employing the algorithm of Kuhn[81] (see Section 3.2.4 and the Supporting Information).

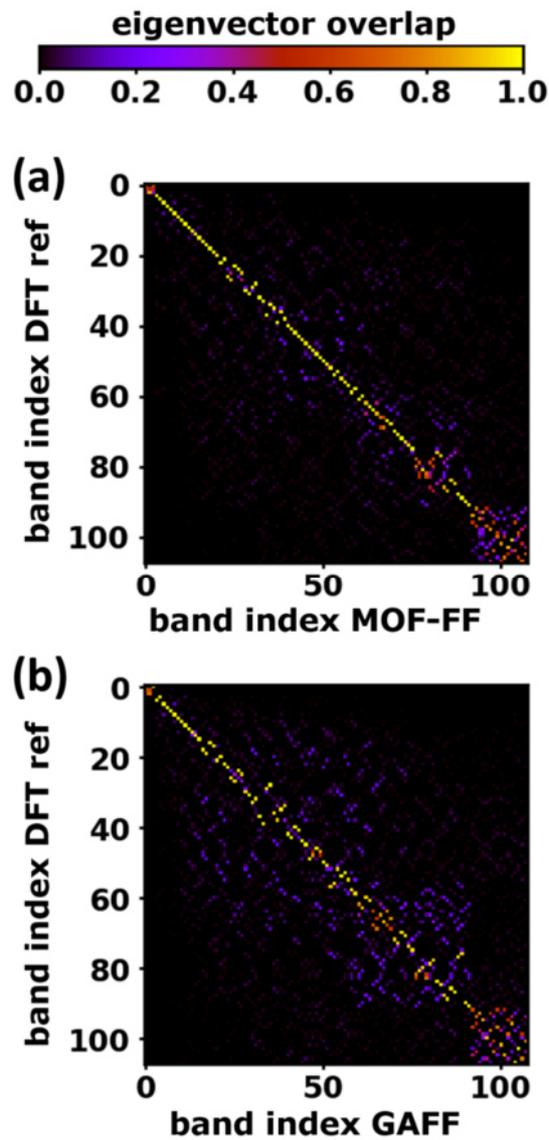

*Figure 13. Overlap matrix of Γ–point eigenvectors of the reference calculation (DFT/D3-BJ) with (a) MOF-FF and (b) GAFF. The overlap matrix $S_{ij}$ is defined as the complex dot product of the $i^{th}$ eigenvector of the reference calculation and the $j^{th}$ eigenvector of the compared calculation.*

A typical outcome of such an analysis is shown in Figure 13 for (a) MOF-FF and (b) GAFF. In MOF-FF the result is rather promising, with the eigenvectors of the $i^{th}$ mode in MOF-FF mostly corresponding to the $i^{th}$ mode in the DFT/D3-BJ reference. I.e., especially for the first ~70 modes the order of the phonon energies is largely preserved, and most MOF-FF modes can be unambiguously associated with a DFT/D3-BJ reference mode. Also the DFTB-based approaches perform rather well in this comparison (see Supporting Information). The good correspondence between approximate modes and reference modes is lost especially in GAFF [see Figure 13(b)] and also in COMPASS (see Supporting Information). For GAFF, starting from the ~20$^{th}$ band, a number of off-diagonal elements of significant magnitude occur. Finally, it should be mentioned that we observe even more significant off-diagonal overlap matrix elements for wavevectors different from Γ, but a detailed analysis of this goes beyond the scope of the current article.

Based on this mode assignment, it is again useful to analyze frequency differences (in analogy to Figure 7) in terms of the RMSD$_f$ values, and average frequency differences AD$_f$. The frequency differences with respect to DFT/D3-BJ and the cumulative RMSD$_f$ values are shown in Figure 14. Additionally,

Table **2** contains the root mean square deviations and the average deviations calculated over the entire frequency range.

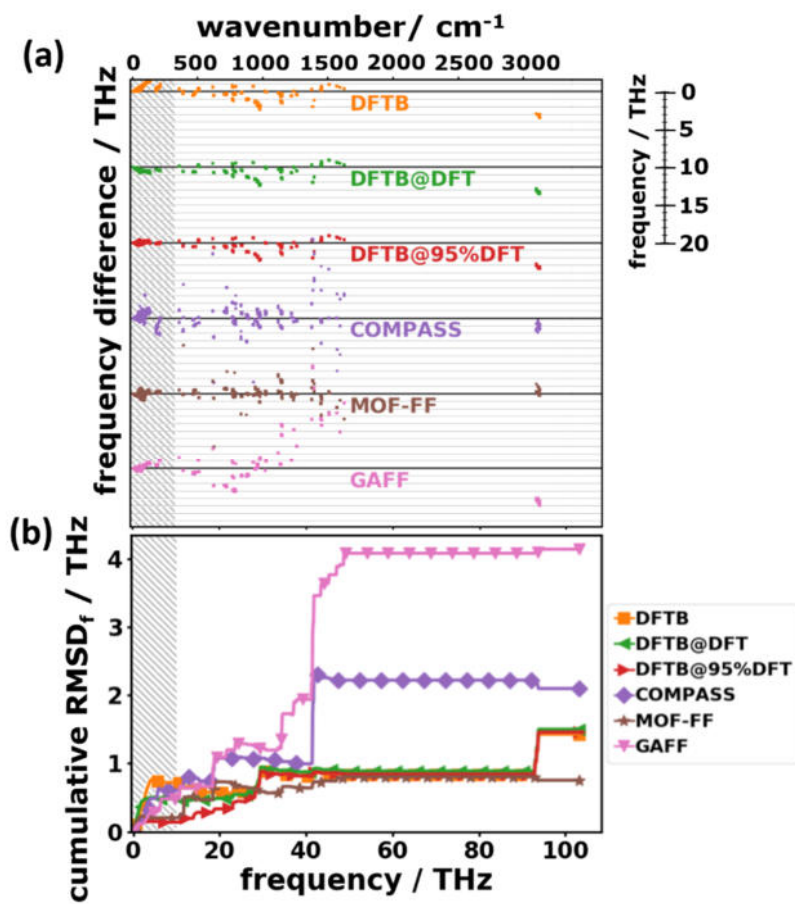

***Figure 14.*** *(a) Frequency differences with respect to the reference data (DFT/D3-BJ) for the various approaches as a function of the reference frequency for the entire phonon spectrum. Each approach has its own zero line (thick black horizontal lines). (b) Cumulative RMSD of frequencies below a certain cutoff frequency as a function of that cutoff frequency. The hatched area highlights the low-frequency regime discussed in Section 3.2. The symbols in (b) do not represent actually calculated data points (these lie much more densely), but rather serve as guides to the eye.*

Figure 14(a) shows that in all cases the observed frequency differences of equivalent modes are much larger than in the low-frequency region. Especially COMPASS and GAFF have massive problems above ~20 THz with large frequency deviations of varying sign, which for some modes amount to more than 10 THz. This results in a steep increase in the $RMSD_f$ for these methods at ~20 THz. The situation deteriorates further for frequencies higher than ~40 THz, where GAFF yields a pronounced overestimation of most phonon frequencies, especially between 35 THz and 50 THz. This can be attributed to harmonic force constants for C-C interactions that are much larger in GAFF than in the DFT/D3-BJ reference (see Supporting Information). Part of the deviations can also be attributed to the neglect of cross-terms in the force field. For COMPASS, too high and too low frequencies are rather equally distributed, resulting in a comparably small value of the $AD_f$, as listed in

Table **2**. An analysis of the displacement patterns reveals that the most affected modes correspond to C-C stretching and C-H in-plane bending motions, whose frequencies often show differences of more than 5 THz, sometimes up to 15 THz. A possible explanation for the poor performance of COMPASS, despite the rather complex nature of the force field, could be the fact that in this approach all atoms of given chemical species are treated equally, which does not very well reflect the actual situation.

MOF-FF outperforms the other force-field based approaches in essentially the entire frequency range with the lowest cumulative $RMSD_f$ values of all tested approaches for frequencies above 30 THz. This is not entirely unexpected, considering that MOF-FF has been specifically parametrized to describe the bonding-type interactions of naphthalene (see Section 2.3), which results in harmonic force constants for the C-C interactions that compare well with the DFT/D3-BJ data (see Supporting Information). Similar to the COMPASS case, there is no systematic over- or underestimation of phonon frequencies.

Outside the low-frequency region, all DFTB-based approaches display a tendency to under- rather than overestimate vibrational frequencies, resulting in negative values of the $AD_f$. For a rather wide spectral range, the DFTB-based results feature an agreement to the reference data as good as MOF-FF. Only around 90 THz the situation deteriorates resulting in nearly twice as high overall RMSD$_f$ values. This is a consequence of the particularly poor description of the C-H stretching vibrations with errors as large as -3 THz, which can be traced back to a pronounced underestimation of the C-H harmonic force constants (see Supporting Information).

### 3.3.3 Analyzing the performance of approximate methods for describing observables derived from the entire phonon spectrum

Base on the above-discussed trends for the calculated phonon frequencies, an analysis of derived thermodynamic properties can be now performed. Regarding the evolution of the heat capacity at temperatures beyond those considered already in Section 3.2.5, Figure 2(b) reveals that at 300 K, the width at half maximum of the low-pass envelope function $f_C$ approaches 20 THz [$f_{1/2}$(300 K) ≈ 18.3 THz], while its tail reaches well beyond 40 THz. Therefore, the frequency differences at higher frequencies that have been discussed in the previous section become increasingly important at higher temperatures: For GAFF, the overestimation of the phonon frequencies between 30 THz and 50 THz results in the distinct underestimation of the heat capacity beyond 200 K shown by the negative values of the errors in the heat capacity, $\Delta C_V$, in Figure 15. In contrast, $\Delta C_V$ remains rather small for COMPASS owing to the fortuitous cancellation of errors that also leads to the virtually vanishing value of the AD$_f$ (see above). MOF-FF displays an excellent performance over essentially the entire frequency range, as here both, the RMSD$_f$, as well as the AD$_f$ adopt very small values. For the DFTB-based approaches, the tendency to mostly underestimate phonon frequencies above ~10 THz results in values of $\Delta C_V$ becoming increasingly positive, where the absolute value of the deviation depends on, whether the low-frequency modes have been over- (DFTB) or also underestimated (DFTB@DFT). Consequently, for DFTB the errors partially cancel, while for DFTB@DFT they add up.

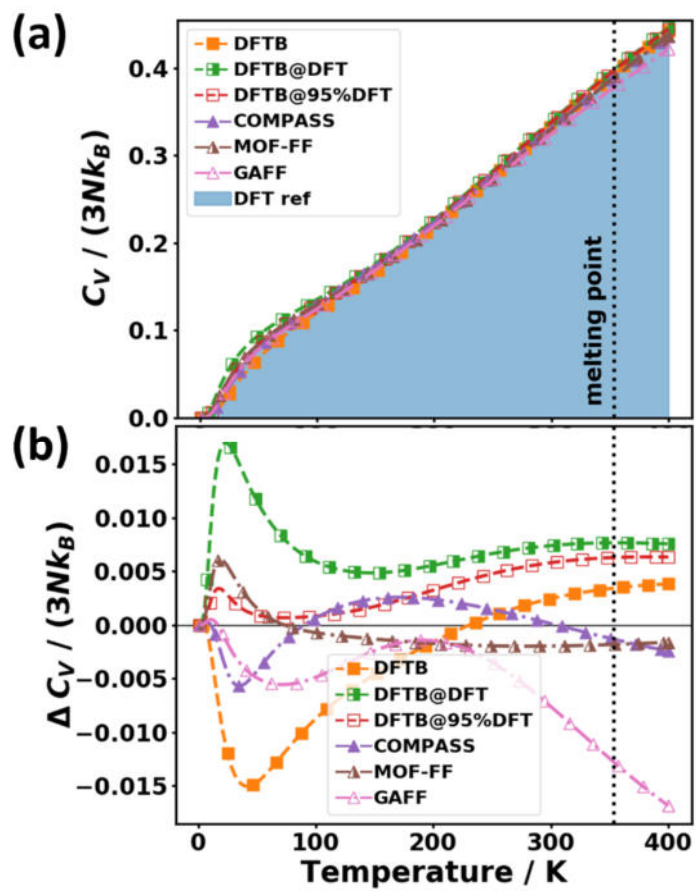

*Figure 15.* (a) Phonon heat capacity $C_V$ normalized by the number of degrees of freedom 3N and the Boltzmann constant $k_B$ as a function of temperature for the tested approaches over a wider temperature range. The shaded area represents the DFT reference. (b) Difference in heat capacity with respect to the reference data (DFT ref). The melting point of the system is indicated by the vertical dotted line[85]. The symbols do not represent actually calculated data points (these lie much more densely), but rather serve as guides to the eye

In spite of the deviations discussed above, it should be mentioned that overall the heat capacity is a rather robust quantity with the relative error at 300 K in none of the cases exceeding 0.5 %.

Another important thermodynamic quantity impacted by all phonons is the Helmholtz free energy $F$. It is the thermodynamic potential for canonical ensembles and is, thus, relevant for thermodynamic stability considerations. The analytic expression[72] for $F$ per unit cell is shown in eq 5

$$F = \frac{1}{N_q}\sum_{\lambda=(n,q)} \left( k_B T \ln\left\{1 - e^{-\frac{\hbar\omega_\lambda}{k_B T}}\right\} + \frac{\hbar\omega_\lambda}{2} \right) \qquad (5),$$

where the first term is a sum over free energy contributions per mode $\lambda$ (characterized by the band index $n$ and the sampled wave vectors $q$) and the second term is the zero-point energy of the harmonic oscillators, defining the free energy at 0 K. $N_q$ denotes the number of wavevectors over which the sampling of the Brillouin zone is carried out. As the zero-point energy contains all harmonic oscillator energies independent of their occupation, the errors in all frequency ranges equally contribute to that quantity. This implies that for the free energy an accurate description of phonons with high frequencies is of distinct relevance, especially at low temperatures, where the zero-point energy dominates. When the temperature increases, the occupation of modes becomes increasingly important, such that then low-frequency modes more strongly impact the temperature dependence of $F$.

Figure 16(a) compares the temperature-dependence of the free energy for all tested approaches. The differences in free energy with respect to the reference calculation are plotted in Figure 16(b). Close to 0 K, one can see that MOF-FF and COMPASS are in closest agreement with the reference. All the DFTB-based approaches yield too small and GAFF too large zero-point energies. This in agreement with the AD$_f$ values listed in Table **2**. COMPASS still fares rather well due to the compensation of errors discussed already in the context of the heat capacity. As far as the absolute magnitude of the error of the free energy is concerned, it can increase beyond 0.25 eV. This value is sizable (amounting to ~10 $k_BT$ at room temperature) and can exceed the energetic differences between different polymorphs typically described in literature[14].

Interestingly, for all force fields the deviations from the reference rather weakly depend on temperature. This also applies to DFTB@95%DFT, which can be attributed to the rather favorable description of the low-frequency modes by this approach (see Section 3.2.2). The situation changes for DFTB and DFTB@DFT. In the former case, the absolute magnitude of the deviation decreases with temperature. This can be explained by the fact that the error in the zero-point energy arises from the predominant underestimation of phonon frequencies when considering the entire frequency range. At higher temperatures, this is increasingly compensated by the overestimation of the energy of the then occupied low-frequency phonons. In contrast, for DFTB@DFT the errors due to the zero-point energy and due to the contribution of low-frequency phonons add up.

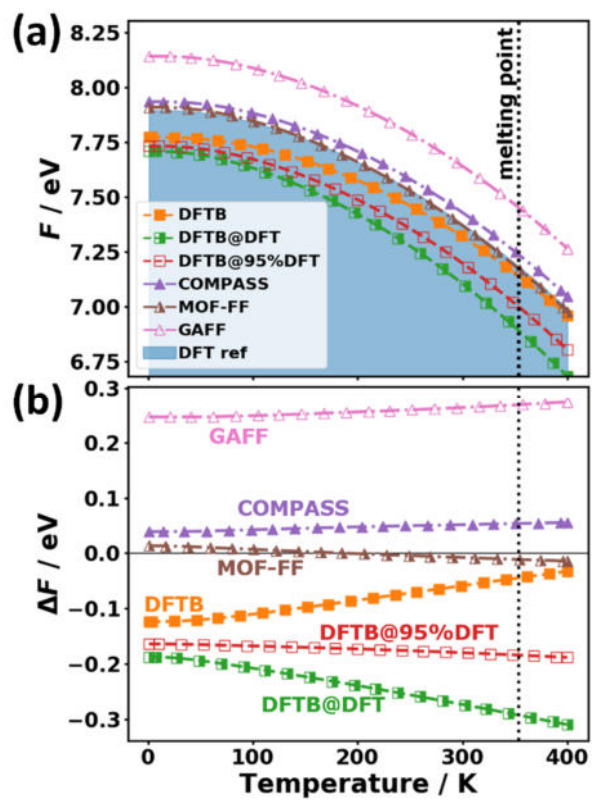

*Figure 16.* (a) Vibrational free energy and (b) difference of free energy with respect to the reference calculations as a function of temperature. The melting point of the system is indicated by the vertical dotted line[85]. The symbols do not represent actually calculated data points (these lie much more densely), but rather serve as guides to the eye.

## 4. SUMMARY AND CONCLUSION

To provide a concise summary of the many aspects discussed above, Figure 17 compares the relative performance of the different approaches for the main quantities of interest. These comprise the frequencies, the group velocities, the mean square thermal displacements, the heat capacities and the free energies. For the quantities for which a statistical analysis is useful, the comparison primarily relies on the root mean square deviations. Only for the frequencies, as the "basic quantities", it also considers the average deviations to get an impression for which methods cancellations of errors could occur, especially when calculating heat capacities or free energies. We compare density-functional tight-binding based and force-field approaches to dispersion-corrected DFT results. For the latter an excellent agreement with measured phonon band structures and Raman spectra can be obtained, provided that a suitable *a posteriori* van der Waals correction is used. As far as the latter is concerned, we observe that the D3-BJ correction clearly outperforms the TS and D2 approaches.

Amongst the approximate methodologies, the tested system-specifically parametrized second-generation force field (MOF-FF) displays clearly the best overall performance. Only in terms of the accuracy of the mean square thermal displacements it is outperformed by the GAFF and COMPASS force fields, owing to their particularly accurate description of the acoustic phonons. This suggests that a strategy for the further improvement of MOF-FF could lie in modifying the way, van der Walls interactions are described in that force field (see Section 2.3). The COMPASS force field fares rather well also in the calculation of heat capacities and free energies in spite of the rather large root mean square deviations for the calculated frequencies. This is a consequence of error cancellations, as COMPASS does not yield a general trend regarding an over- or underestimation of frequencies, which is consistent with very small deviations for the average frequencies. In contrast, GAFF rather overestimates frequencies with the effect being particularly pronounced in the intermediate frequency range for intramolecular vibrations (see Figure 14). This results in the tendency to underestimate the heat capacity (as modes become occupied only at higher temperatures). Consistently, GAFF overestimates free energies in the entire spectral range. The DFTB-based approaches, despite the considerably increase computational costs typically perform worse than the force fields, especially worse than MOF-FF. The only exception is the rather accurate description of phonon frequencies in the low-frequency region in the case of DFTB@95%DFT, which can be attributed to the tuning of the unit-cell size in this approach. The tendency of the DFTB-based approaches to rather underestimate frequencies of intramolecular vibrations results in a distinct underestimation of the zero-point energy. For the free energy at room temperature, DFTB benefits from error cancellations due to an overestimation of the frequencies of the increasingly occupied phonons in the low-frequency region.

Notably, Figure 17 shows relative deviations, i.e., deviations compared to the worst performing methodology. Thus, it is also worthwhile to separately address the general performance of the approximate methodologies for the different phonon-related quantities of interest: for example, the description for the heat capacity is rather satisfactory for all approaches, not exceeding 0.5% at room temperature even for the worst performing method. The errors for the free energy are larger reaching ~3 % for several of the applied methodologies. At room temperature, the observed deviations of the free energy of up to ~10 $k_BT$ per unit cell, in fact, exceed commonly observed energy differences between different organic polymorphs[10,12–14]. In this context it should be noted that especially MOF-FF with its comparably accurate description of phonon frequencies does not show this problem and yields free energies within ~±0.01 eV compared to the PBE/D3-BJ values over a wide temperature range. The observables most sensitive to the applied methodology are the mean square thermal displacements, which are overestimated by a factor of two by the DFTB@DFT calculations and where only the COMPASS and GAFF force fields display a satisfactory performance. The reason for that is that this quantity is primarily determined by the properties of the acoustic phonons, whose frequencies are so low that even minor absolute errors of the calculated frequencies result in major relative errors and, thus, an incorrect description of the thermal motion of the atoms.

These considerations show that approximate methodologies for describing phonon bands in organic semiconductors are promising for the description of phonon-related properties in molecular crystals at affordable computational cost. Especially, system-specifically parametrized force fields have a high potential. Nevertheless, there is still quite some room for improvements, where our results suggest that especially advancing the description of the van der Waals interactions in the tested system-specific force field should be a promising avenue.

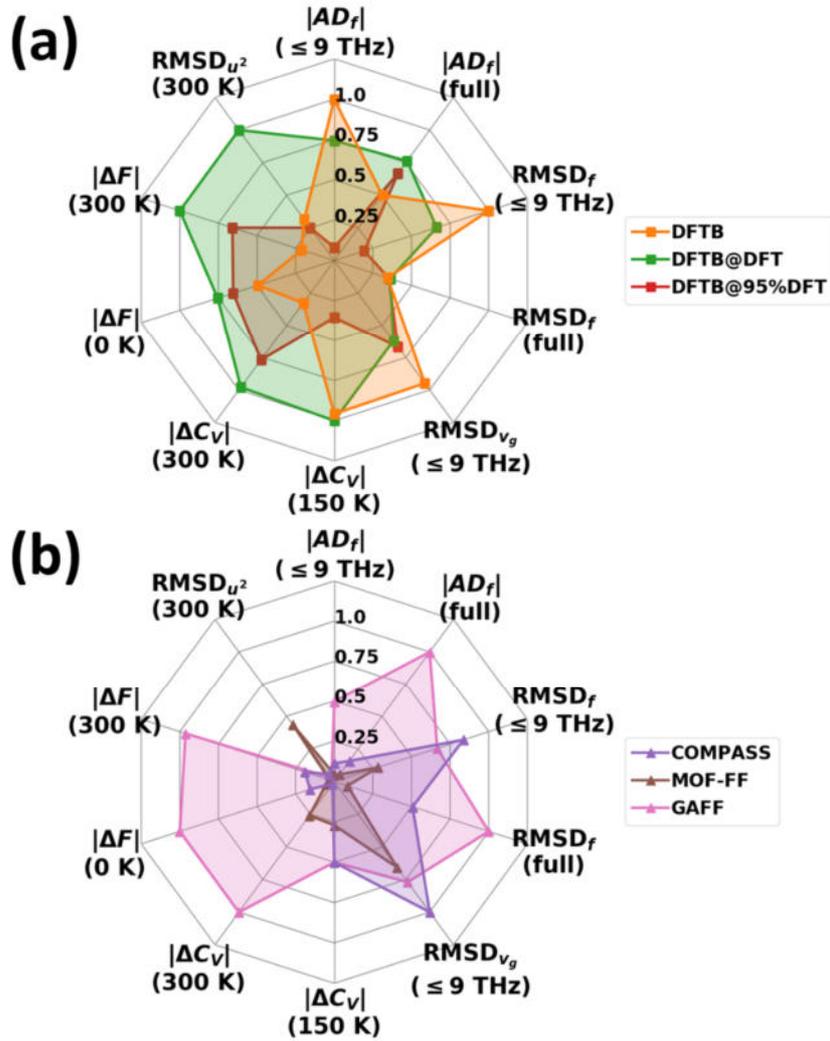

*Figure 17.* Radar charts summarizing the performance of (a) DFTB-, and (b) FF-based methodologies with respect to the reference calculations. The category axes are normalized such that 1 corresponds to the maximum observed error indicator amongst all approaches. The compared quantities comprise the average deviations in the calculated frequencies with respect to the DFT/D3-BJ reference, $AD_f$, for the low-frequency region ($\leq 9$ THz) and the full spectral range (full), the root mean square deviations of frequencies ($RMSD_f$) for the low-frequency region ($\leq 9$ THz) and the full spectral range (full). Additionally, the RMSD value of norms of group velocity vectors stemming from phonon modes in the low-frequency region ($RMSD_{v_g}$), and the absolute

*differences in thermodynamic properties (Helmholtz free energy F and heat capacity $C_V$) at various temperatures are shown. Finally, the root mean square deviation of mean squared thermal displacements ($RMSD_{u2}$) is shown, which is calculated as the average (quadratic) deviation of thermal displacements summed over all atoms in the unit cell.*

ASSOCIATED CONTENT

**Supporting Information**

The Supporting Information is available free of charge on the ACS Publications website at DOI:

Further computational details including the effect of numerical parameters, the methodology to simulate Raman spectra, the FF parameters, further comparisons to experimental data including tests employing the MBD correction for Γ-phonons, the assignment procedure of phonon modes, an analysis of the harmonic force constants, and thermodynamic properties at very high temperatures. (PDF)

AUTHOR INFORMATION

**Corresponding Author**

Egbert Zojer – *Institute of Solid State Physics, Graz University of Technology, Petersgasse 16, 8010 Graz, Austria*; Email: egbert.zojer@tugraz.at

**Notes**

The authors declare no competing financial interest.

ACKNOWLEDGEMENTS

T.K., S.W. and E.Z. acknowledge the Graz University of Technology for financial support through the Lead Project (LP-03) and the use of HPC resources provided by the ZID. The computational results have been in part achieved by using the Vienna Scientific Cluster (VSC3). H.K. acknowledges the financial support through the JST CREST (Grant Number JPMJCR1813), Japan, and the kind support by Prof. M. Nakamura at the Nara Institute of Science and Technology (NAIST), Japan. N.B. gratefully acknowledges financial support under the COMET program within the K2 Center "Integrated Computational Material, Process, and Product Engineering (IC-MPPE)" (Project No. 859480). This program is supported by the Austrian Federal Ministries for Transport, Innovation, and Technology (BMVIT) and for Digital and Economic Affairs (BMDW), represented by the Austrian research funding association (FFG), and the federal states of Styria, Upper Austria, and Tyrol. This work has partially funded by the Austrian Climate and Energy Fund (KLIEN) and the Austrian Research Promotion Agency (FFG) in the course through the "ThermOLED" project (FFG No. 848905). J.P.D. and R.S. acknowledge financial support from the Deutsche Forschungsgemeinschaft (DFG, Grants: SCHM 1389/10-1 within FOR 2433 and SCHM 1389/8-1)

ABBREVIATIONS

1BZ, first Brillouin zone; RMSD, root mean square deviation; AD, average difference; DFT, density functional theory; DFTB, density functional tight binding; FF, force field; MBD, many-body dispersion; DOS, density of states; DOGV, density of group velocities; vdW, van der Waals; MSTD, mean squared thermal displacement; FWHM, full width at half maximum.

# SUPPORTING INFORMATION

# Evaluating Computational Shortcuts in Supercell-Based Phonon Calculations of Molecular Crystals: The Instructive Case of Naphthalene

*Tomas Kamencek[1,2], Sandro Wieser[1], Hirotaka Kojima[3], Natalia Bedoya-Martínez[4], Johannes P. Dürholt[5], Rochus Schmid[5], and Egbert Zojer[1]*

[1]*Institute of Solid State Physics, Graz University of Technology, NAWI Graz, Petersgasse 16, 8010 Graz, Austria*

[2]*Institute of Physical and Theoretical Chemistry, Graz University of Technology, NAWI Graz, Stremayrgasse 9, 8010 Graz, Austria*

[3]*Division of Materials Science, Nara Institute of Science and Technology, 8916-5 Takayama, Ikoma, Nara 630-0192, Japan*

[4]*Materials Center Leoben, Roseggerstraße 12, 8700 Leoben, Austria*

[5]*Chair of Inorganic Chemistry 2, CMC Group, Ruhr University Bochum, Universitätsstraße 150, 44801 Bochum, Germany*

1. **Importance of sampling phonons in the entire reciprocal space for calculating thermodynamic properties**

In spite of the practice common in literature to neglect phonon dispersions when calculating thermodynamic properties, non-Γ-phonons can have a significant impact on the results. Fig. S1 shows how the evolutions of Helmholtz free energy and the heat capacity differ, when considering only Γ-phonons or phonons from the entire first Brillouin zone (sampled on a 9×10×9 mesh; see main text). For the free energy one observes an energy difference of more than 0.1 eV per unit cell at room temperature. Contrary, the heat capacity is mainly influenced at low temperatures: if only Γ-phonons are considered, there is a non-vanishing contribution at zero frequency giving rise to a violation of the third law of thermodynamics.

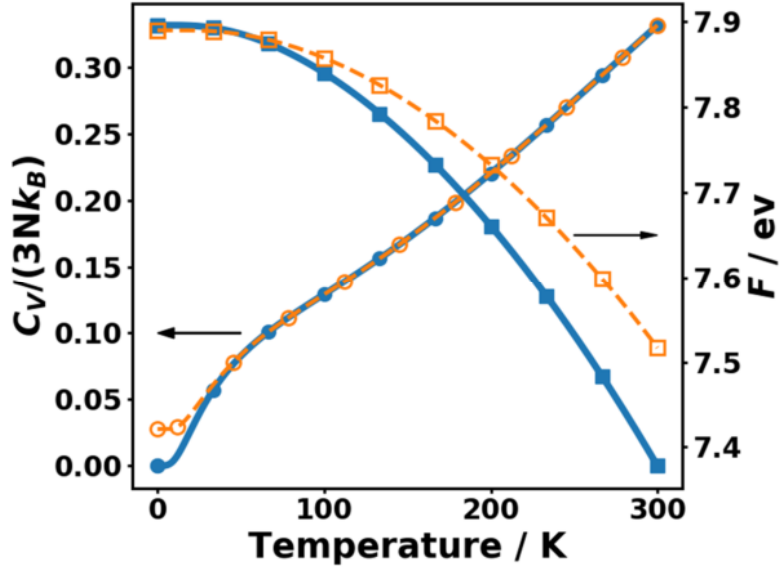

**Fig. S1:** *Phonon heat capacity $C_V$ normalized by the number of vibrational modes 3N (circles) and vibrational free energy F (squares) as a function of temperature calculated from phonons of the entire first Brillouin zone (blue solid lines) and from Γ-phonons only (orange dashed lines).*

## 2. Raman spectra simulation

The intensities $I_k$ associated with the $k^{th}$ vibrational mode is related to the Raman activity $A_k$ according to the following equations [1]. Besides the Raman activity, the calculated (Stokes) intensity depends on the fourth power of the frequency difference between the vibrational mode and the excitation radiation as well as a thermal occupation factor including the Bose-Einstein distribution $n$ (eq. (1a)).

The Raman activity $A_k$ depends on the geometry of the experimental setup and on the sample. This is considered in eq. (1b) for the assumption of the incident and the reflected light beam being orthogonal to each other. For isotropic and homogeneous samples, the Raman activity can be written according to eq. (1b) with the symbols being defined in eq. (1c) and (1d).

$$I_k \propto \frac{(\Omega_{in} - \omega)^4}{\omega_k} \ (n(\omega, T) + 1) \ A_k \tag{1a}$$

$$A_k = 45 \gamma_k^2 + 7 \beta_k^2 \tag{1b}$$

$$\gamma_k = \frac{1}{3} \sum_{i=1}^{3} \chi_{ii,k} \tag{1c}$$

$$\beta_k^2 = \frac{1}{2}\{(\chi_{11,k} - \chi_{22,k})^2 + (\chi_{11,k} - \chi_{33,k})^2 + (\chi_{33,k} - \chi_{22,k})^2 \quad (1d)$$
$$+ 6(\chi_{12,k}^2 + \chi_{23,k}^2 + \chi_{13,k}^2)\}$$

$$\chi_{ij,k} = \frac{\partial \alpha_{ij}}{\partial Q_k} = \sum_{l=1}^{3N} \frac{\partial \alpha_{ij}}{\partial u_l} \frac{e_k^{(l)}}{\sqrt{m_l}} \quad (1e)$$

The most relevant quantities appearing in these expressions are the derivatives of the polarization tensor $\alpha_{ij}$ with respect to normal mode coordinates $Q_k$. They consititute the Raman tensor, $\chi_{ij,k}$. The Raman tensor can either be calculated directly by displacing the geometry along the normal mode coordinates and calculating the change in polarizability (or electric susceptibility) as a function of the normal mode displacement.

Alternatively, one can rewrite the derivative with respect to the normal mode coordinate $Q_k$ as a derivative with respect to the cartesian displacement $u_l$. The associated transformation matrix is then given by the (mass weighted) phonon eigenvectors (polarization vectors) $e_k^{(l)}$. Although the direct approach is more useful when calculating Raman tensors for specific modes, the approach based on cartesian displacements is much more efficient for systems with large number of symmetries since many cartesian derivatives can be obtained from simple symmetry transformations, and, thus, the number of symmetry-inequivalent displacements necessary to simulate Raman tensors for all modes can be drastically reduced.

Practically, the dielectric function was calculated with *VASP* applying density functional perturbation theory for each displaced geometry produced by *PHONOPY*. The symmetry-irreducible Raman tensors were calculated from eq. (1e), while the remaining ones were obtained by applying the respective point group symmetry operations to those rank 3 tensors.

The Raman activities of the isolated naphthalene molecule were calculated with the *Gaussian 16* package (Revision A.03) [2] after a proper geometry optimization employing the D3-BJ van-der-Waals *a posteriori* correction. For both functionals used (PBE and B3LYP) we employed the triple-zeta Gaussian-type basis set 6-311++G(d,p) including diffuse and polarization functions. Subsequently, the Raman activities (using the calculation type identifiers `Opt` and `Freq`) were calculated using the equations above with the fully automatic routines of *Gaussian*.

The plotted spectra consist of Lorentzian function with a full-width-at-half-maximum of 0.2 THz placed at each resonance.

## 3. Converging DFT settings

### 3.1 Used Pseudopotentials and global setting

The following *VASP* pseudopotentials were used for hydrogen and carbon, respectively: PAW_PBE H 15Jun2001, PAW_PBE C 08Apr2002

Additionally, the following simulation settings were used throughout all tests:

```
LREAL =.FALSE.;
ALGO = Fast;
ISMEAR = 0;
SIGMA = 0.05;
```

### 3.1 Impact of DFT settings on phonon frequencies

Especially the low frequency phonon bands are often found to be relatively sensitive to the simulation parameters, so that tight convergence criteria must be chosen, consuming a high amount of computational resources. We, thus, studied how the *VASP*-specific parameters controlling the plane wave energy cutoff (`ENCUT`), the SCF convergence criterion (`EDIFF`) and a global precision setting (`PREC`) influence the resulting Γ frequencies. The geometry, which was optimized with the thoroughly converged parameters described in the main text, was kept the same for all tests. The root-mean-square (RMS) error and the maximum absolute deviation of the Γ frequencies with respect to the reference simulation were recorded. The associated maps can be seen Fig. S2. We find that the errors depend much more strongly on the energy cutoff than on the SCF convergence criterion. Furthermore, the error does not monotonically decrease with the cutoff. Interestingly, for smaller cutoffs, choosing normal precision results in lower RMS errors than the accurate setting. The supposedly most underconverged settings (top left in Fig. S2) also result in smaller RMS errors than calculations with the same energy cutoff but different SCF convergence criteria

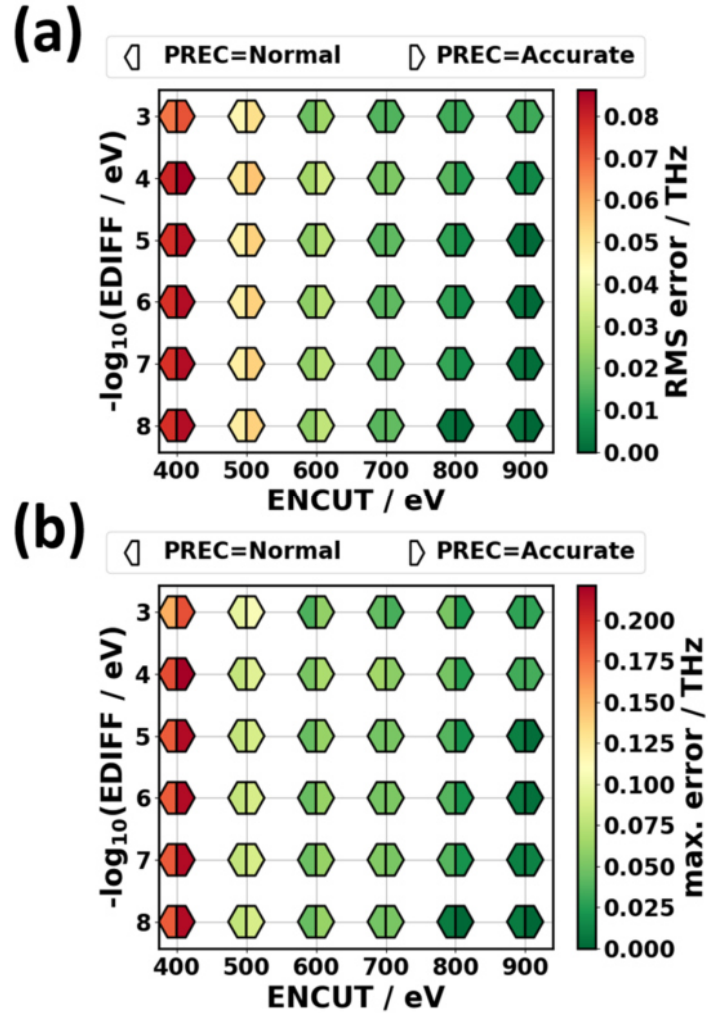

**Fig. S2:** *Accuracy map of Γ frequencies dependent on the three VASP parameters `PREC`, `ENCUT`, `EDIFF`, controlling the global precision settings, the plane wave energy cutoff, and the SCF convergence criterion, respectively. The (a) root-mean-square (RMS) error and the (b) maximum error were calculated with respect to the reference calculation (`PREC` =Accurate, `ENCUT` =900 eV, `ENCUT` =$10^{-8}$ eV).*

Although the gain in computing time for the most underconverged settings is not extra-ordinarily large (about a factor 3.7 given our computing architecture Intel Xeon E5-25650 CPUs, the used level of parallelization and our compilation of the code), it is still instructive to compare the phonon bands obtained with the most "economic" settings to the most accurate reference results. We base the comparison of the band structure on the low frequency modes (below 9 THz ≈ 300 cm$^{-1}$) due to the reasons given in the Methodology section of the main manuscript. The band structure and the density of states (DOS) obtained with the "economic" settings are compared to the reference calculation in Fig. S3. Although the results are supposedly highly underconverged, the agreement in the bands and especially in the DOS is surprisingly good. Only the width of the band gap between ~4.1-5.3 THz slightly increases, as the bands at the lower edge of the gap are somewhat shifted to lower frequencies At ~2 THz and ~3 THz at the A point some further small differences can be seen. Notably, the lowest acoustic band (transverse acoustic) in ΓA direction is the only acoustic band, whose band width is notably underestimated. The most pronounced difference is the band dispersion of the second lowest band along XA which is much flatter in "DFT eco" than in the reference calculation "DFT ref".

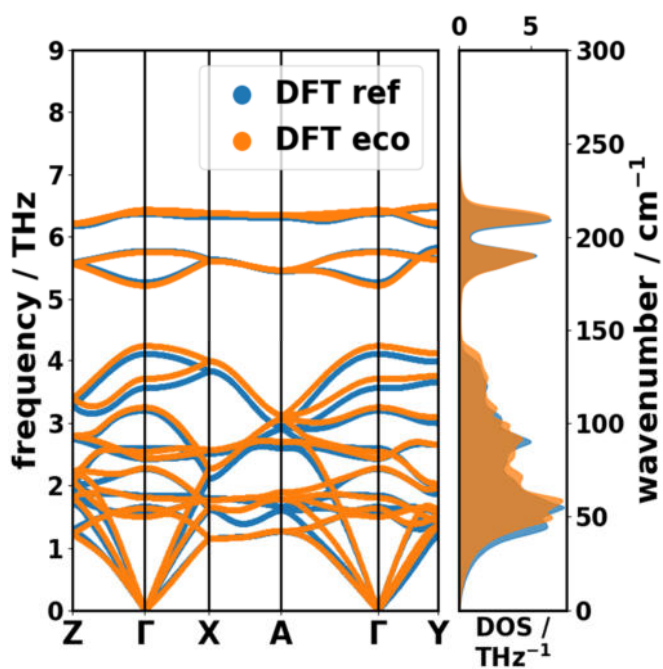

**Fig. S3:** *Phonon band structure and DOS of (non-deuterated) naphthalene obtained with DFT+D3-BJ. Blue: accurate reference calculation. Orange: underconverged ("economic") DFT settings*

To quantify the deviation, we sampled the 1BZ at 125 **q**-points and calculated the RMS deviation between the "economic" and the reference data. When considering all phonons of the material, an RMS deviation of 0.08 THz is obtained, which ich only slightly larger than the calculated RMS deviation, when considering only Γ-point frequencies (0.07 THz). The RMS deviation for bands up to frequencies of 9 THz is slightly increased (0.11 THz), which suggests that the reduced simulation accuracy more severely impacts the low frequency modes of mostly inter-molecular character. The higher-lying, intra-molecular modes are apparently less affected.

### 4. D3-BJ parameters used in DFTB and DFT

The D3-BJ van der Waals (vdW) correction depends on four global parameters ($a_1$, $a_2$, $s_6$, and $s_8$) with $s_6$ usually being kept fixed at unity. The other three parameters were chosen according to the suggested standard values provided in the respective user manuals [3],[4]. In this work, we used the recommended standard parameters for the PBE and DFTB3 functional, respectively.

**Tab. S1:** *Standard D3-BJ parameters used for the given functionals according to the VASP and DFTB+ manuals.*

| Functional | a$_1$ / Bohr | a$_2$ | s$_6$ | s$_8$ |
|---|---|---|---|---|
| DFTB3 | 0.746 | 4.191 | 1.0 | 3.209 |
| PBE | 0.4289 | 4.4407 | 1.0 | 0.7875 |

## 5. Cell optimization and unit cell rescaling in DFTB

Unlike *VASP*, which allows to optimize unit cell parameters within the constraint of constant volume to perform a fit to an equation of state, the used version of *DFTB+* (version 18.1) does not provide this functionality. To overcome this problem, a different approach was chosen for optimizing the unit cell while keeping the type of Bravais lattice (simple monoclinic lattice with four lattice parameters: the three lengths of lattice vectors *a*, *b*, and *c* as well as the monoclinic angle β): for a set of fixed monoclinic angles, the lattice vector lengths (together with the atomic coordinates) were optimized with fixed angles. Afterwards the optimum monoclinic angle was obtained by fitting a parabola to the energy-vs.-angle data (see Fig. S4). This optimum angle was used in a last step for a unit cell, whose remaining lattice vectors and atomic positions were optimized to end up with the final fully optimized geometry.

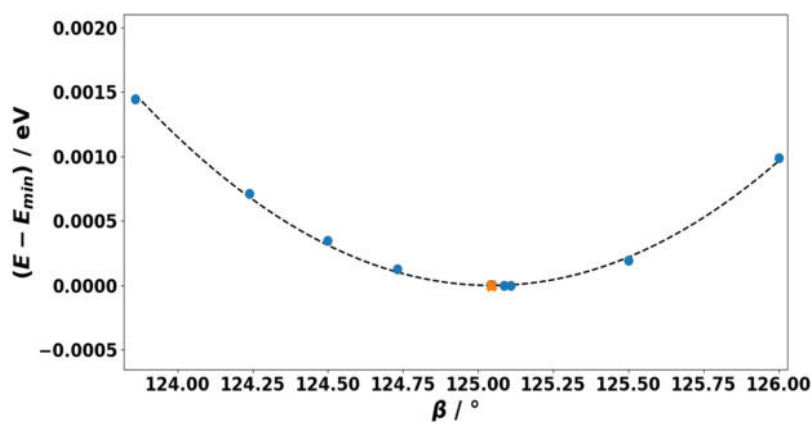

**Fig. S4:** *Total energy difference with respect to the minimum as a function of the monoclinic angle β calculated with DFTB+. The blue circles correspond to optimized unit cells with fixed monoclinic angle, the orange cross marks the fitted minimum.*

## 6. Unit cell rescaling in DFTB

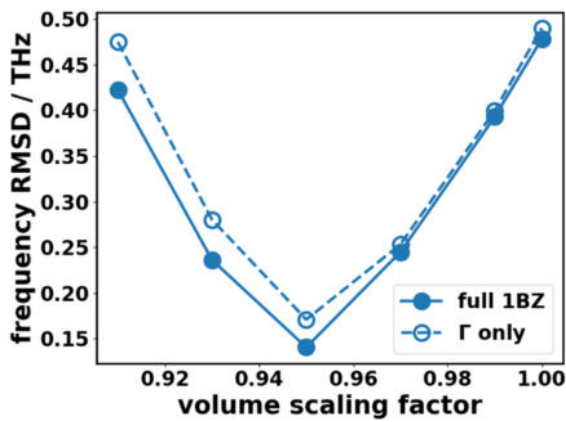

**Figure S5.** *Root-mean-square deviation (RMSD) of frequencies between DFTB calculations and the DFT/D3-BJ reference. The DFTB frequencies have been obtained for a DFT/D3-BJ unit cell, whose volume has been rescaled by the factor given on the horizontal axis (for details see main text). The solid line with filled symbols represents the deviations calculated for the entire first Brillouin zone, while for the dashed line with empty symbols $\Gamma$–point frequencies have been considered. Note that the DFTB energy of the "optimally" rescaled unit cell is ~30 meV higher than for the DFTB-optimized cell and shows a tensile stress of ~ -5 kbar.*

## 7. MOF-FF parametrization

The MOF-FF functional form is built as a sum over many independent contributions [5]:

$$E = E_{str} + E_{bnd} + E_{tor} + E_{oop} + E_{coul} + E_{vdW} \qquad (2a)$$

Here $E_{str}$ is the stretch bond potential, $E_{bnd}$ is the bending potential, $E_{tor}$ is the torsional potential, $E_{oop}$ is the out of plane bending potential, $E_{coul}$ is the electrostatic potential and $E_{vdW}$ is the van der Waals potential.

**Bonded interaction terms:**

$$E_s^{str} = \frac{1}{2} k_s (r_s - r_s^{ref})^2 \left[ 1 - 2.55(r_s - r_s^{ref}) + \frac{7}{12}\left(2.55\left(r_s - r_s^{ref}\right)\right)^2 \right] \qquad (2b)$$

$$E_b^{bnd} = \frac{1}{2}k_b(\theta - \theta_b^{ref})^2 \left[1 - 0.014(\theta_b - \theta_b^{ref}) + 5.6 \cdot 10^{-5}(\theta_b - \theta_b^{ref})^2 - 7 \right. \tag{2c}$$
$$\left. \cdot 10^{-7}(\theta_b - \theta_b^{ref})^3 + 2.2 \cdot 10^{-8}(\theta_b - \theta_b^{ref})^4\right]$$

$$E_t^{tor} = \sum_n \frac{V_t^n}{2}[1 + \cos(n\phi_t + \phi_t^n)] \tag{2d}$$

$$E_o^{oop} = \frac{1}{2}k_o(\theta_o - \theta_0)^2 \tag{2e}$$

**Cross-terms between bonded interactions between atoms of a bending angle:**

$$E_b^{str-bnd} = (\theta_b - \theta_b^{ref})[k_{sb1}(r_{b1} - r_{b1}^{ref}) + k_{sb2}(r_{b2} - r_{b2}^{ref})] \tag{2f}$$

$$E_b^{str-str} = k_{ss}(r_{b1} - r_{b1}^{ref})(r_{b2} - r_{b2}^{ref}) \tag{2g}$$

**Torsional cross-term, that was not originally contained in MOF-FF (underlined in the following input file):**

$$E_{bb13} = N(r_{ij} - r_1)(r_{kl} - r_3) \tag{2h}$$

**Non-bonded interaction terms:**

$$E_{ij}^{vdW} = \varepsilon_{ij}\left\{1.85 \cdot 10^5 \exp\left(-12\frac{d_{ij}}{d_{ij}^0}\right) - 2.25\left(\frac{d_{ij}^0}{d_{ij}}\right)^6 \left[1 + 6\left(\frac{0.25 d_{ij}^0}{d_{ij}}\right)^{14}\right]^{-1}\right\} \tag{2i}$$

$$E_{ij}^{coul} = \frac{1}{4\pi\varepsilon_0}\frac{q_1 q_2}{d_{ij}}\mathrm{erf}\left(\frac{d_{ij}}{\sigma_{ij}}\right) \tag{2j}$$

$k, \alpha, V$ ... fitted parameters

$r$ ... interatomic distance

$\theta$ ... bending angle

$\phi$ ... torsional angle

$d_{ij}$ ... non-bonded atomic distance

$\epsilon_{ij}$ ... van der Waals potential well depth

$q_i$ ... atomic point charge

$\sigma_{ij}$ ... Gaussian charge distribution width

The reference data obtained with the Turbomole [6] software package (version 7.3) was obtained utilizing the PBE functional [7], employing the D3-BJ [8],[9] dispersion correction and using the def2-TZVPP [10] basis set. The SCF convergence criterion was set to $10^{-7}$ Hartree (~$2 \cdot 10^{-6}$ eV). The atomic position was optimized until a maximum force of $10^{-3}$ Hartree/Bohr (~$5 \cdot 10^{-2}$ eV/ Å) and an energy convergence of $10^{-6}$ Hartree (~$2 \cdot 10^{-5}$ eV) was reached.

The parameters for the force field potential used for obtaining vibrational properties can be found in form of a LAMMPS input file. The syntax of the parameters can be found in the official manual: https://lammps.sandia.gov/doc/Manual.html

```
# MOF-FF Potential parameters
# Explanation of atom descriptors:
# c3_c2h1: atom of species c bonded to three different atoms,
#          two of which are of species c and one of species h
# c3_c2h1S: the S indicates the atoms closer to the molecule center
pair_style buck6d/coul/gauss/long 0.9000    0.9000    12.0000

pair_coeff     1    1         10304        3.0612245       457.17971       4.5218516
0.60800971    # buck6d->(c3_c2h1@naph)|naphthalene/gaussian->(c3_c2h1@naph)|naphthalene <-->
buck6d->(c3_c2h1@naph)|naphthalene/gaussian->(c3_c2h1@naph)|naphthalene

pair_coeff     1    2         10304        3.0612245       457.17971       4.5218516
0.60800971    # buck6d->(c3_c2h1@naph)|naphthalene/gaussian->(c3_c2h1@naph)|naphthalene <-->
buck6d->(c3_c2h1S@naph)|naphthalene/gaussian->(c3_c2h1S@naph)|naphthalene

pair_coeff     1    3         10304        3.0612245       457.17971       4.5218516
0.60800971    # buck6d->(c3_c2h1@naph)|naphthalene/gaussian->(c3_c2h1@naph)|naphthalene <-->
buck6d->(c3_c3@naph)|naphthalene/gaussian->(c3_c3@naph)|naphthalene

pair_coeff     1    4        6157.8178      3.4682081       129.19572       0.78772886
0.73006542    # buck6d->(c3_c2h1@naph)|naphthalene/gaussian->(c3_c2h1@naph)|naphthalene <-->
buck6d->(h1_c1@naph)|naphthalene/gaussian->(h1_c1@naph)|naphthalene

pair_coeff     1    5        6157.8178      3.4682081       129.19572       0.78772886
0.73006542    # buck6d->(c3_c2h1@naph)|naphthalene/gaussian->(c3_c2h1@naph)|naphthalene <-->
buck6d->(h1_c1S@naph)|naphthalene/gaussian->(h1_c1S@naph)|naphthalene

pair_coeff     2    2         10304        3.0612245       457.17971       4.5218516
0.60800971    # buck6d->(c3_c2h1S@naph)|naphthalene/gaussian->(c3_c2h1S@naph)|naphthalene <-->
buck6d->(c3_c2h1S@naph)|naphthalene/gaussian->(c3_c2h1S@naph)|naphthalene

pair_coeff     2    3         10304        3.0612245       457.17971       4.5218516
0.60800971    # buck6d->(c3_c2h1S@naph)|naphthalene/gaussian->(c3_c2h1S@naph)|naphthalene <-->
buck6d->(c3_c3@naph)|naphthalene/gaussian->(c3_c3@naph)|naphthalene
```

```
pair_coeff     2     4        6157.8178      3.4682081       129.19572      0.78772886
0.73006542    # buck6d->(c3_c2h1S@naph)|naphthalene/gaussian->(c3_c2h1S@naph)|naphthalene <-->
buck6d->(h1_c1@naph)|naphthalene/gaussian->(h1_c1@naph)|naphthalene

pair_coeff     2     5        6157.8178      3.4682081       129.19572      0.78772886
0.73006542    # buck6d->(c3_c2h1S@naph)|naphthalene/gaussian->(c3_c2h1S@naph)|naphthalene <-->
buck6d->(h1_c1S@naph)|naphthalene/gaussian->(h1_c1S@naph)|naphthalene

pair_coeff     3     3          10304        3.0612245       457.17971      4.5218516
0.60800971    # buck6d->(c3_c3@naph)|naphthalene/gaussian->(c3_c3@naph)|naphthalene <-->
buck6d->(c3_c3@naph)|naphthalene/gaussian->(c3_c3@naph)|naphthalene

pair_coeff     3     4        6157.8178      3.4682081       129.19572      0.78772886
0.73006542    # buck6d->(c3_c3@naph)|naphthalene/gaussian->(c3_c3@naph)|naphthalene <-->
buck6d->(h1_c1@naph)|naphthalene/gaussian->(h1_c1@naph)|naphthalene

pair_coeff     3     5        6157.8178      3.4682081       129.19572      0.78772886
0.73006542    # buck6d->(c3_c3@naph)|naphthalene/gaussian->(c3_c3@naph)|naphthalene <-->
buck6d->(h1_c1S@naph)|naphthalene/gaussian->(h1_c1S@naph)|naphthalene

pair_coeff     4     4           3680            4            32.805       0.10690769
0.9771554     # buck6d->(h1_c1@naph)|naphthalene/gaussian->(h1_c1@naph)|naphthalene <-->
buck6d->(h1_c1@naph)|naphthalene/gaussian->(h1_c1@naph)|naphthalene

pair_coeff     4     5           3680            4            32.805       0.10690769
0.9771554     # buck6d->(h1_c1@naph)|naphthalene/gaussian->(h1_c1@naph)|naphthalene <-->
buck6d->(h1_c1S@naph)|naphthalene/gaussian->(h1_c1S@naph)|naphthalene

pair_coeff     5     5           3680            4            32.805       0.10690769
0.9771554     # buck6d->(h1_c1S@naph)|naphthalene/gaussian->(h1_c1S@naph)|naphthalene <-->
buck6d->(h1_c1S@naph)|naphthalene/gaussian->(h1_c1S@naph)|naphthalene

bond_style hybrid class2 morse harmonic

bond_coeff     3 class2    1.096263    366.570728    -934.755355   1390.448591    # mm3-
>(c3_c2h1@naph,h1_c1@naph)|naphthalene

bond_coeff     6 class2    1.433161    381.108715    -971.827224   1445.592996    # mm3-
>(c3_c3@naph,c3_c3@naph)|naphthalene

bond_coeff     1 class2    1.391210    490.285107   -1250.227022   1859.712695    # mm3-
>(c3_c2h1@naph,c3_c2h1S@naph)|naphthalene

bond_coeff     2 class2    1.454449    349.932367    -892.327535   1327.337208    # mm3-
>(c3_c2h1@naph,c3_c2h1@naph)|naphthalene

bond_coeff     4 class2    1.435896    383.965503    -979.112033   1456.429149    # mm3-
>(c3_c2h1S@naph,c3_c3@naph)|naphthalene

bond_coeff     5 class2    1.090635    382.608333    -975.651250   1451.281234    # mm3-
>(c3_c2h1S@naph,h1_c1S@naph)|naphthalene
```

```
angle_style hybrid class2/p6 cosine/buck6d
```

```
angle_coeff      8 class2/p6       108.146325    94.230110   -75.585826    17.322995   -12.406682    22.341015    # mm3->(c3_c2h1S@naph,c3_c3@naph,c3_c3@naph)|naphthalene

angle_coeff      8 class2/p6 bb    61.790580     1.435896    1.433161

angle_coeff      8 class2/p6 ba    69.044622    24.678564    1.435896    1.433161

angle_coeff      7 class2/p6       120.083549   50.819349   -40.764279    9.342485   -6.691062    12.048759    # mm3->(c3_c2h1S@naph,c3_c3@naph,c3_c2h1S@naph)|naphthalene

angle_coeff      7 class2/p6 bb    63.409397     1.435896    1.435896

angle_coeff      7 class2/p6 ba    33.560423    33.560423    1.435896    1.435896

angle_coeff      6 class2/p6       120.989674   30.577529   -24.527487    5.621286   -4.025950     7.249626    # mm3->(c3_c3@naph,c3_c2h1S@naph,h1_c1S@naph)|naphthalene

angle_coeff      6 class2/p6 bb    7.682251      1.435896    1.090635

angle_coeff      6 class2/p6 ba    25.498143    28.515360    1.435896    1.090635

angle_coeff      3 class2/p6       116.807583   39.917331   -32.019324    7.338289   -5.255662     9.463999    # mm3->(c3_c2h1@naph,c3_c2h1@naph,h1_c1@naph)|naphthalene

angle_coeff      3 class2/p6 bb    8.763731      1.454449    1.096263

angle_coeff      3 class2/p6 ba    30.724541    26.104046    1.454449    1.096263

angle_coeff      4 class2/p6       114.769310   93.785723   -75.229365   17.241300   -12.348172    22.235656    # mm3->(c3_c2h1@naph,c3_c2h1S@naph,c3_c3@naph)|naphthalene

angle_coeff      4 class2/p6 bb    90.168492     1.391210    1.435896

angle_coeff      4 class2/p6 ba    86.536187    63.628816    1.391210    1.435896

angle_coeff      5 class2/p6       124.234272   29.012370   -23.272009    5.333552   -3.819875     6.878542    # mm3->(c3_c2h1@naph,c3_c2h1S@naph,h1_c1S@naph)|naphthalene

angle_coeff      5 class2/p6 bb    9.732514      1.391210    1.090635

angle_coeff      5 class2/p6 ba    31.449745    26.623502    1.391210    1.090635

angle_coeff      1 class2/p6       112.413065   99.632083   -79.918970   18.316079   -13.117925    23.621769    # mm3->(c3_c2h1@naph,c3_c2h1@naph,c3_c2h1S@naph)|naphthalene

angle_coeff      1 class2/p6 bb    104.912889    1.454449    1.391210

angle_coeff      1 class2/p6 ba    106.286485   77.649009    1.454449    1.391210

angle_coeff      2 class2/p6       118.001113   38.562917   -30.932893    7.089297   -5.077335     9.142881    # mm3->(c3_c2h1S@naph,c3_c2h1@naph,h1_c1@naph)|naphthalene
```

```
angle_coeff     2 class2/p6 bb     9.862886    1.391210    1.096263

angle_coeff     2 class2/p6 ba     31.790211   26.090601   1.391210    1.096263

dihedral_style hybrid opls class2

dihedral_coeff    4 class2     0.000000    0.000000    2.279729    0.000000    0.000000    0.000000    # class2->(h1_c1@naph,c3_c2h1@naph,c3_c2h1S@naph,h1_c1S@naph)|naphthalene

dihedral_coeff    4 class2 mbt 0.0 0.0 0.0 0.0

dihedral_coeff    4 class2 ebt 0.0 0.0 0.0 0.0 0.0 0.0 0.0 0.0

dihedral_coeff    4 class2 at  0.0 0.0 0.0 0.0 0.0 0.0 0.0 0.0

dihedral_coeff    4 class2 aat 0.0 0.0 0.0

dihedral_coeff    4 class2 bb13 0.0 0.0 0.0

dihedral_coeff   11 class2     0.000000    0.000000    2.415742    0.000000    0.000000    0.000000    # class2->(h1_c1S@naph,c3_c2h1S@naph,c3_c3@naph,c3_c3@naph)|naphthalene

dihedral_coeff   11 class2 mbt 0.0 0.0 0.0 0.0

dihedral_coeff   11 class2 ebt 0.0 0.0 0.0 0.0 0.0 0.0 0.0 0.0

dihedral_coeff   11 class2 at  0.0 0.0 0.0 0.0 0.0 0.0 0.0 0.0

dihedral_coeff   11 class2 aat 0.0 0.0 0.0

dihedral_coeff   11 class2 bb13 0.0 0.0 0.0

dihedral_coeff    8 class2     0.000000    0.000000    1.897513    0.000000    0.000000    0.000000    # class2->(c3_c2h1@naph,c3_c2h1S@naph,c3_c3@naph,c3_c2h1S@naph)|naphthalene

dihedral_coeff    8 class2 mbt 0.0 0.0 0.0 0.0

dihedral_coeff    8 class2 ebt 0.0 0.0 0.0 0.0 0.0 0.0 0.0 0.0

dihedral_coeff    8 class2 at  0.0 0.0 0.0 0.0 0.0 0.0 0.0 0.0

dihedral_coeff    8 class2 aat 0.0 0.0 0.0

dihedral_coeff    8 class2 bb13 0.0 0.0 0.0

dihedral_coeff    2 class2     0.000000    0.000000    3.996343    0.000000    0.000000    0.000000    # class2->(c3_c2h1@naph,c3_c2h1@naph,c3_c2h1S@naph,h1_c1S@naph)|naphthalene

dihedral_coeff    2 class2 mbt 0.0 0.0 0.0 0.0

dihedral_coeff    2 class2 ebt 0.0 0.0 0.0 0.0 0.0 0.0 0.0 0.0

dihedral_coeff    2 class2 at  0.0 0.0 0.0 0.0 0.0 0.0 0.0 0.0
```

```
dihedral_coeff     2 class2 aat 0.0 0.0 0.0

dihedral_coeff     2 class2 bb13 0.0 0.0 0.0

dihedral_coeff    10 class2    0.000000    0.000000    1.639432    0.000000    0.000000    0.000000    # class2->(h1_c1S@naph,c3_c2h1S@naph,c3_c3@naph,c3_c2h1S@naph)|naphthalene

dihedral_coeff    10 class2 mbt 0.0 0.0 0.0 0.0

dihedral_coeff    10 class2 ebt 0.0 0.0 0.0 0.0 0.0 0.0 0.0 0.0

dihedral_coeff    10 class2 at  0.0 0.0 0.0 0.0 0.0 0.0 0.0 0.0

dihedral_coeff    10 class2 aat 0.0 0.0 0.0

dihedral_coeff    10 class2 bb13 0.0 0.0 0.0

dihedral_coeff    12 class2    0.000000    0.000000    2.799249    0.000000    0.000000    0.000000    # class2->(c3_c2h1S@naph,c3_c3@naph,c3_c3@naph,c3_c2h1S@naph)|naphthalene

dihedral_coeff    12 class2 mbt 0.0 0.0 0.0 0.0

dihedral_coeff    12 class2 ebt 0.0 0.0 0.0 0.0 0.0 0.0 0.0 0.0

dihedral_coeff    12 class2 at  0.0 0.0 0.0 0.0 0.0 0.0 0.0 0.0

dihedral_coeff    12 class2 aat 0.0 0.0 0.0

dihedral_coeff    12 class2 bb13 0.0 0.0 0.0

dihedral_coeff     1 class2 bb13   -70.099189     1.454449     1.435896    # bb13->(c3_c2h1@naph,c3_c2h1@naph,c3_c2h1S@naph,c3_c3@naph)|naphthalene

dihedral_coeff     1 class2    0.000000    0.000000    3.801176    0.000000    0.000000    0.000000    # class2->(c3_c2h1@naph,c3_c2h1@naph,c3_c2h1S@naph,c3_c3@naph)|naphthalene

dihedral_coeff     1 class2 mbt 0.0 0.0 0.0 0.0

dihedral_coeff     1 class2 ebt 0.0 0.0 0.0 0.0 0.0 0.0 0.0 0.0

dihedral_coeff     1 class2 at  0.0 0.0 0.0 0.0 0.0 0.0 0.0 0.0

dihedral_coeff     1 class2 aat 0.0 0.0 0.0

dihedral_coeff     3 class2    0.000000    0.000000    4.025101    0.000000    0.000000    0.000000    # class2->(h1_c1@naph,c3_c2h1@naph,c3_c2h1S@naph,c3_c3@naph)|naphthalene

dihedral_coeff     3 class2 mbt 0.0 0.0 0.0 0.0

dihedral_coeff     3 class2 ebt 0.0 0.0 0.0 0.0 0.0 0.0 0.0 0.0

dihedral_coeff     3 class2 at  0.0 0.0 0.0 0.0 0.0 0.0 0.0 0.0

dihedral_coeff     3 class2 aat 0.0 0.0 0.0

dihedral_coeff     3 class2 bb13 0.0 0.0 0.0
```

```
dihedral_coeff     9 class2     0.000000     0.000000     7.137224     0.000000     0.000000     0.000000     # class2->(c3_c2h1@naph,c3_c2h1S@naph,c3_c3@naph,c3_c3@naph)|naphthalene

dihedral_coeff     9 class2 mbt 0.0 0.0 0.0 0.0

dihedral_coeff     9 class2 ebt 0.0 0.0 0.0 0.0 0.0 0.0 0.0 0.0

dihedral_coeff     9 class2 at  0.0 0.0 0.0 0.0 0.0 0.0 0.0 0.0

dihedral_coeff     9 class2 aat 0.0 0.0 0.0

dihedral_coeff     9 class2 bb13 0.0 0.0 0.0

dihedral_coeff     6 class2     0.000000     0.000000     2.585528     0.000000     0.000000     0.000000     # class2->(c3_c2h1S@naph,c3_c2h1@naph,c3_c2h1@naph,h1_c1@naph)|naphthalene

dihedral_coeff     6 class2 mbt 0.0 0.0 0.0 0.0

dihedral_coeff     6 class2 ebt 0.0 0.0 0.0 0.0 0.0 0.0 0.0 0.0

dihedral_coeff     6 class2 at  0.0 0.0 0.0 0.0 0.0 0.0 0.0 0.0

dihedral_coeff     6 class2 aat 0.0 0.0 0.0

dihedral_coeff     6 class2 bb13 0.0 0.0 0.0

dihedral_coeff     5 class2 bb13    -75.085553     1.391210     1.391210     # bb13->(c3_c2h1S@naph,c3_c2h1@naph,c3_c2h1@naph,c3_c2h1S@naph)|naphthalene

dihedral_coeff     5 class2     0.000000     0.000000     6.412985     0.000000     0.000000     0.000000     # class2->(c3_c2h1S@naph,c3_c2h1@naph,c3_c2h1@naph,c3_c2h1S@naph)|naphthalene

dihedral_coeff     5 class2 mbt 0.0 0.0 0.0 0.0

dihedral_coeff     5 class2 ebt 0.0 0.0 0.0 0.0 0.0 0.0 0.0 0.0

dihedral_coeff     5 class2 at  0.0 0.0 0.0 0.0 0.0 0.0 0.0 0.0

dihedral_coeff     5 class2 aat 0.0 0.0 0.0

dihedral_coeff     7 class2     0.000000     0.000000     1.336357     0.000000     0.000000     0.000000     # class2->(h1_c1@naph,c3_c2h1@naph,c3_c2h1@naph,h1_c1@naph)|naphthalene

dihedral_coeff     7 class2 mbt 0.0 0.0 0.0 0.0

dihedral_coeff     7 class2 ebt 0.0 0.0 0.0 0.0 0.0 0.0 0.0 0.0

dihedral_coeff     7 class2 at  0.0 0.0 0.0 0.0 0.0 0.0 0.0 0.0

dihedral_coeff     7 class2 aat 0.0 0.0 0.0

dihedral_coeff     7 class2 bb13 0.0 0.0 0.0

improper_style inversion/harmonic
```

```
improper_coeff    2    1.217508    0.000000    # harm-
>(c3_c2h1S@naph,c3_c2h1@naph,c3_c3@naph,h1_c1S@naph)|naphthalene

improper_coeff    1    5.430825    0.000000    # harm-
>(c3_c2h1@naph,c3_c2h1@naph,c3_c2h1S@naph,h1_c1@naph)|naphthalene

improper_coeff    3    7.389044    0.000000    # harm-
>(c3_c3@naph,c3_c2h1S@naph,c3_c2h1S@naph,c3_c3@naph)|naphthalene
```

**special_bonds lj 0.00 0.00 1.00 coul 1.00 1.00 1.00**

## 8. Super cell convergence

In order to converge the dynamical matrix, one must increase the size of the considered supercell until the obtained change in frequencies becomes negligible. Fig. S6 shows the supercell convergence behavior of the PBE/D3-BJ settings. While the 1×2×1 is still much too small, starting from a 2×2×2 supercell, all displayed bands already have the right dispersion. Only in the ΓA direction, the acoustic bands display slight differences compared to the bands of larger supercells. The chosen 2×3×2 supercell shows virtually no difference to the 3×3×3 supercell but comes at much lower cost. For this reason, the 2×3×2 supercell was considered for the reference DFT calculation. The reason why the chosen number of unit cells in the *b*-direction is larger (3 instead of 2) is to account for the shorter lattice constant in that direction (see main text). This results in a similar "probing radius" for the interatomic force constants in all three spatial directions, while keeping the total system size at a level small enough.

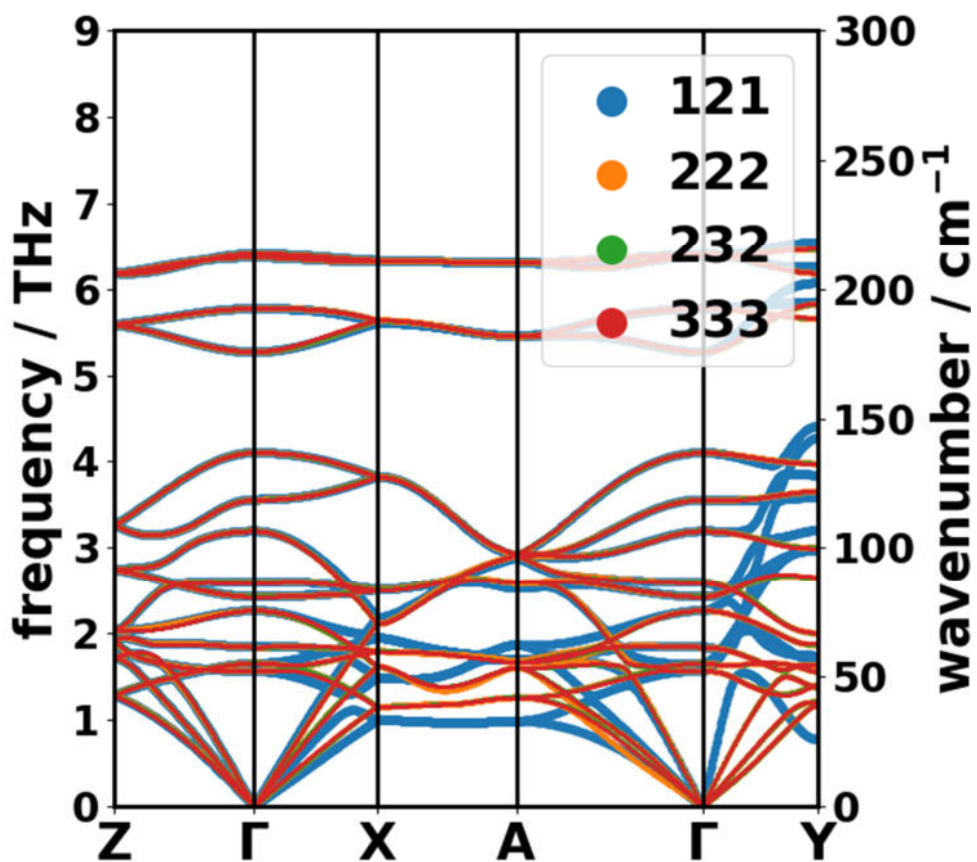

**Fig. S6:** *Convergence of phonon band structure (obtained with PBE/D3-BJ) with respect to supercell size.*

For the force field calculations, 3×3×3 supercells were used. As it is shown in Fig. S7 for the example of MOF-FF, this supercell is a good compromise between accuracy and effort, as there is no significant difference to a 4×4×4 supercell, while the 2×2×2 supercell shows minimal differences in ΓX direction.

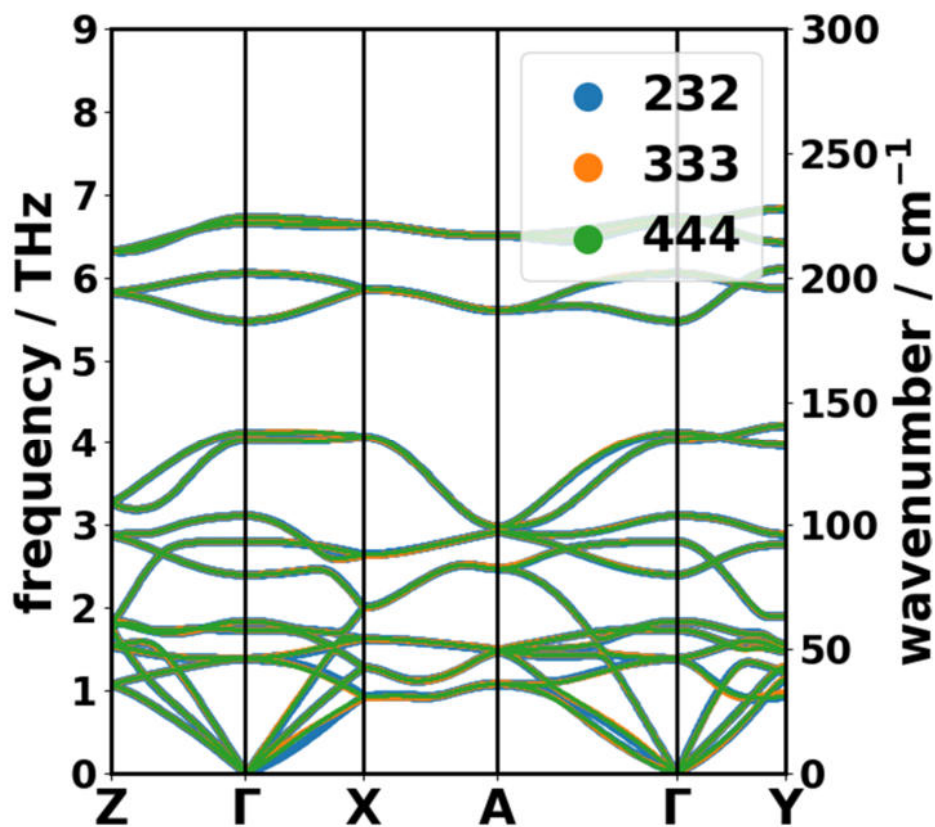

**Fig. S7:** *Convergence of phonon band structure (obtained with MOF-FF) with respect to supercell size.*

## 9. Comparison of Γ-frequencies for different van der Waals corrections

In the context of testing different van der Waals (vdW) corrections, the D3-BJ [8],[9], MBD [11],[12], D2 [13] and TS [14] schemes were used to fully optimize atomic coordinates and unit cells, and compute phonon frequencies. Since we have not been able to simulate phonon bands with the MBD correction, we compare the Γ-frequencies of the four approaches with the (few) experimental data points at this point in reciprocal space. This comparison in Fig. S8 shows that the MBD and the D3-BJ approach essentially yield the same phonon frequencies, while the other two approaches differ quite significantly in the low-frequency region (< 8 THz). As this frequency region is governed by intermolecular motion, the vdW interaction is especially important in this frequency regime.

At higher frequencies, the four approaches are in better agreement, although D2 tends to slightly underestimate frequencies above ~40 THz.

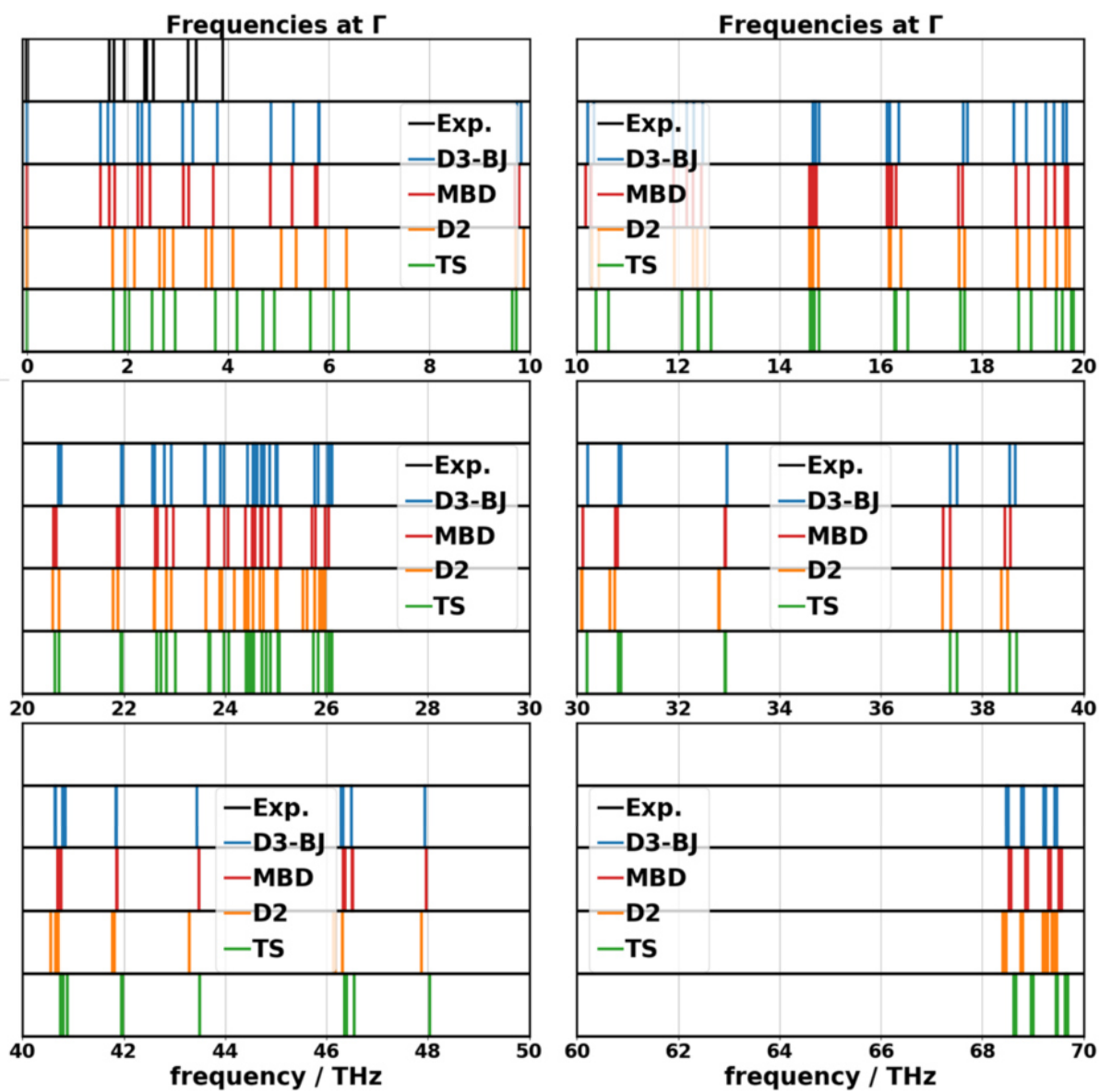

**Fig. S8:** *Comparison of Γ-frequencies obtained with different vdW corrections. The unit cells were fully relaxed employing the same correction.*

## 10. Comparison of phonon band structure of the deuterated systems to experiment

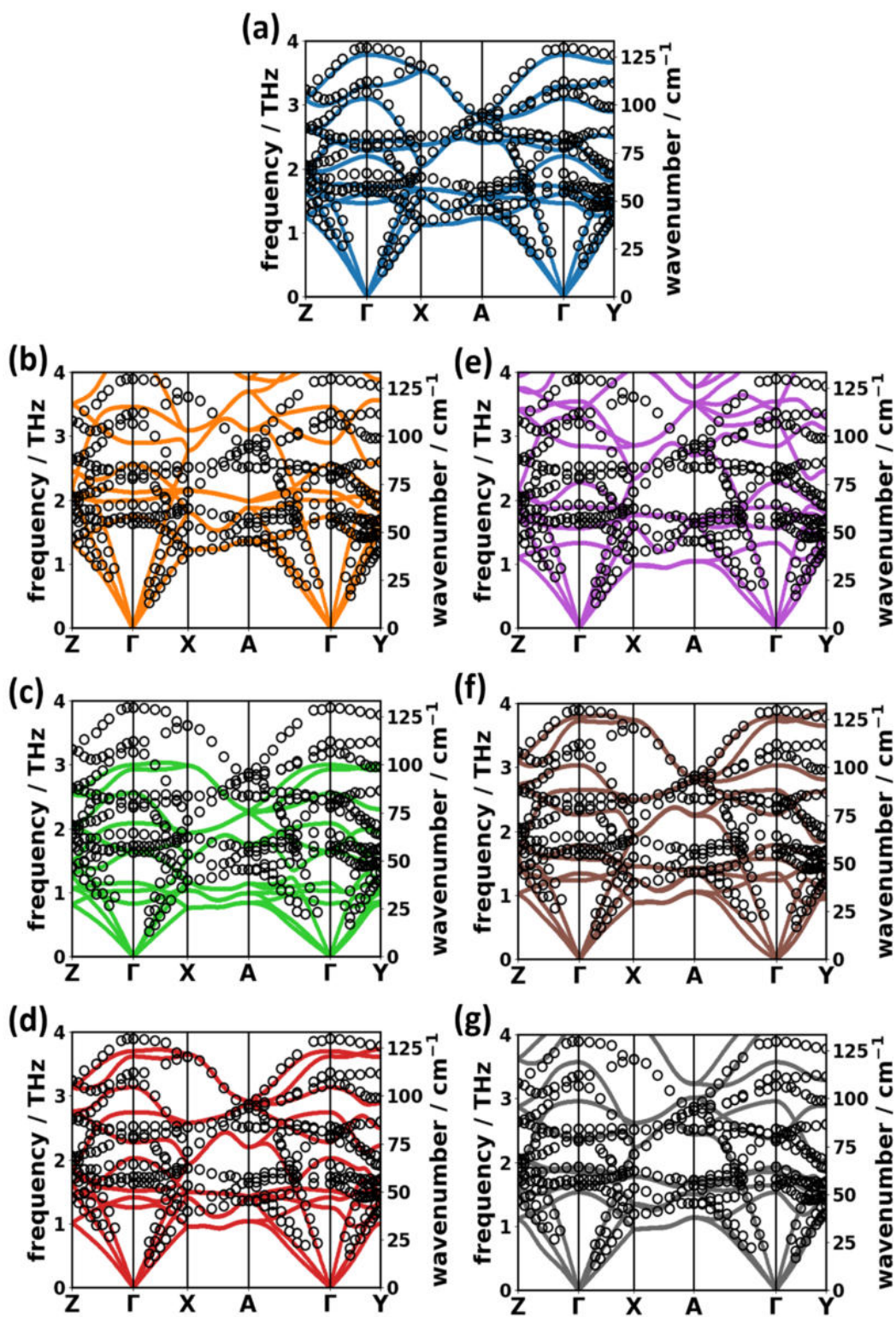

**Fig. S9:** *Phonon band structures (solid lines) of deuterated naphthalene obtained with (a) the DFT reference, (b) DFTB, (c) DFTB@DFT, (d) DFTB@95%DFT, (e) COMPASS, (f) MOF-FF, and (g) GAFF compared to experimental data points (open circles).*

As shown in Fig. S9, all trends discussed in the main manuscript for the comparison of approximate methods to DFT-D3BJ reference data are recovered also for the comparison to the experiments.

## 11. Assignment of vibrational modes

The so-called "Hungarian algorithm" of Kuhn [15] was used to solve the linear assignment problem – i.e. solving the problem of assigning each element of a set an element of a second set minimizing a cost function in that way. In our case, we have two sets of frequencies (phonons) stemming from two different methodology, which we want to assign based on the similarity of their eigenvectors. In that problem the expression (3a) is minimized, with $C_{ij}$ being a cost function and $X$ being a matrix, whose entry $X_{ij}$ is 1 if element $i$ is assigned to element $j$. Here, $X_{ij}$ is 1 if the $i^{th}$ mode in the reference is assigned to the $j^{th}$ mode of the comparison.

$$\sum_{ij} C_{ij} X_{ij} \quad (3a)$$

$$C_{ij} = (1 - S_{ij}) + P_{ij} \quad (3b)$$

$$S_{ij} = e_{ref,i}^{\dagger} \cdot e_{compared,j} \quad (3c)$$

$$P_{ij} = A\left(1 - \exp\left\{-\frac{(\omega_{ref,i} - \omega_{compared,j})^2}{2\sigma^2}\right\}\right) \quad (3d)$$

The cost function must be adapted to the specific problem. Here, we define the cost function [see eq. (3b)] to consist of a matrix $S_{ij}$ which characterizes the eigenvector overlap (complex dot product defined in eq. (3c); $S_{ij}$ = 1 for perfect agreement, $S_{ij}$ =0 for orthogonal eigenvectors) of the $i^{th}$ eigenvector in the reference with the $j^{th}$ eigenvector of the comparison. Additionally, we add a penalty function $P_{ij}$ of Gaussian shape additionally penalizing mode assignments with large frequency differences [see eq. (3d)]. The parameters to choose are the amplitude $A$ of the penalty and the Gaussian width σ. For the reported mode assignments, we used $A$=0.5 and σ=1 THz.

The penalty matrices, the eigenvector overlap and the resulting cost function for the mode assignment at the Γ-point are shown in Fig. S10. Note that especially COMPASS and GAFF tend to show larger off-diagonal elements of the eigenvector overlap matrices, implying that the order of the eigenmodes changes compared to the DFT reference. Moreover, especially with these methods we observe several matrix entries significantly deviating from 1 and 0 indicating that the mode assignment is no longer unique. All tested methodologies show the biggest discrepancies in the nearly degenerate C-H stretching vibrations at ~90 THz.

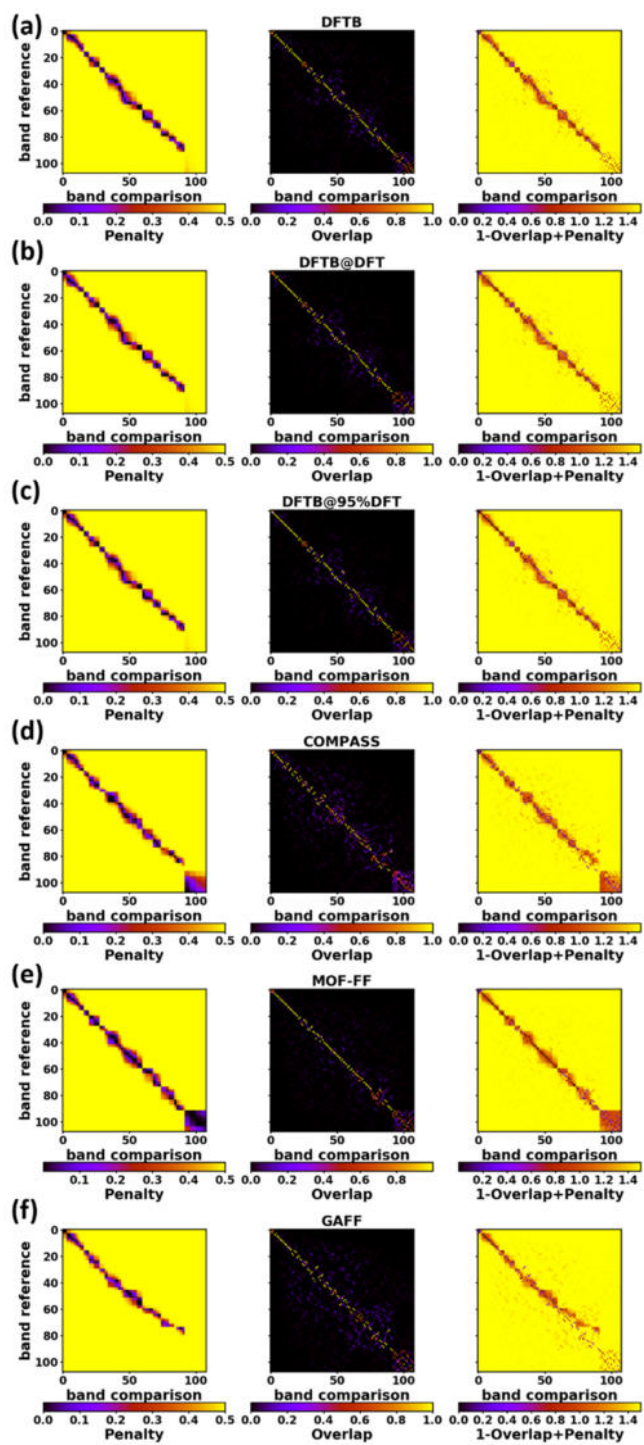

**Fig. S10:** *Penalty (left), overlap (center), and cost (right) functions of the (a) DFTB, (b) DFTB@DFT, (c) DFTB@95%, (d) COMPASS, (e) MOF-FF, and (f) GAFF phonon modes at Γ compared to DFT ref.*

## 12. Raman spectrum of molecular naphthalene using the B3LYP hybrid functional

In order to be consistent with the periodic DFT calculations, the molecular Raman spectra has first been calculated with the same functional. Additionally, the hybrid functional B3LYP [16],[17] has been used to obtain a comparison to a an approach more commonly applied when considering molecular systems [keeping the 6-311G(d,p)++ basis set and the D3-BJ van der Waals correction]. Fig. S11 compares the experimental data of Zhao and McCreery [18] with the crystal Raman spectrum (PBE/D3-BJ) and the molecular spectra (PBE/D3-BJ and B3LYP/D3-BJ). Notably, for the isolated molecule, the PBE simulation is in better agreement with the measured data than the spectrum calculated with the hybrid functional B3LYP, which slightly overestimates the vibrational frequencies. Moreover, the deviation between the B3LYP and PBE spectra is larger than the deviation between the PBE crystal and the PBE molecule simulation. It should be stressed that no empirical frequency scaling factors have been used.

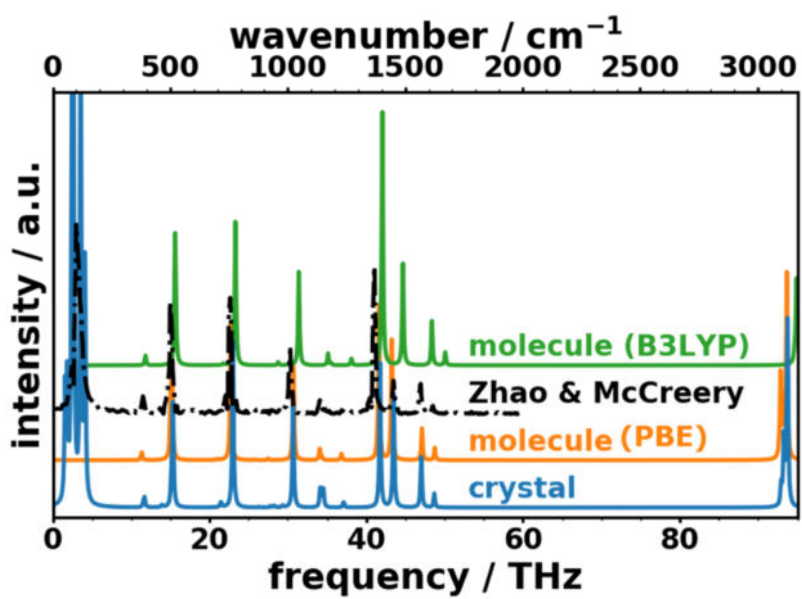

**Fig. S11:** *Simulated unpolarized Raman spectra for molecular (calculated with Gaussian16, 6-311G(d,p)++/PBE and 6-311G(d,p)/B3LYP) and crystalline (VASP, PBE) naphthalene (solid lines) compared to experimental data from Zhao and McCreery [18]. For both, the simulated spectra and the measurement of Zhao and McCreery, an excitation wavelength of 784 nm was used.*

## 13. Analysis of harmonic force constants

In order to track the origin of the most pronounced discrepancies in the phonon spectra, we will briefly comment on the relations of the observed (qualitative) discrepancies in the phonon band structures and the quantities that were actually calculated in the different approaches, namely the interatomic harmonic force constants (HFC). As the latter are rank 2 tensors, we base our discussion on the (rotational invariant) traces of the tensors to arrive at a qualitative, direction independent measure.

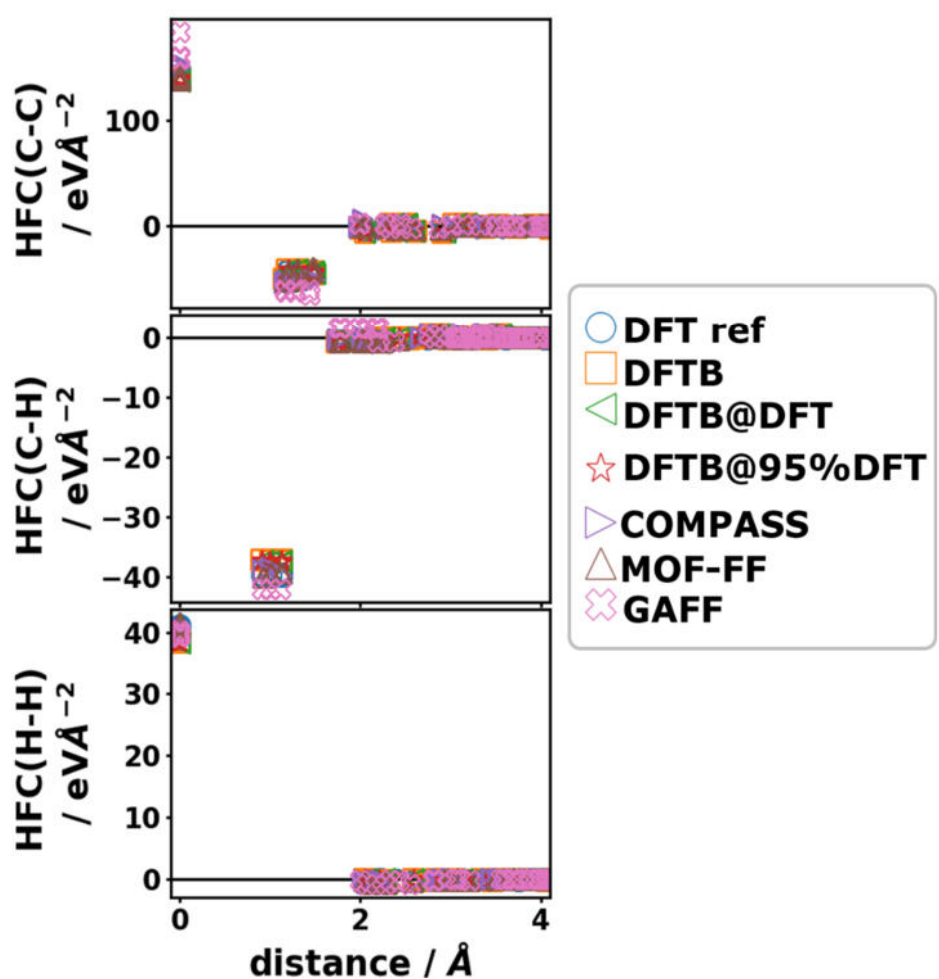

**Fig. S12:** *Trace of harmonic force constant tensors (HFCs) as a function of the distance between the involved atoms (considering periodic boundary conditions).*

The traces of the HFC tensors (in the following referred to as the HFCs) decrease very rapidly to zero (within ~3 Å) with the pair distance of the involved atoms (considering periodic boundary conditions) as shown in Fig. S12. This finding implies that all relevant interactions can be seized within the spatial extents dictated by the considered supercell. Obviously, the HFCs are largest for such atom pairs which are covalently bonded (C-C and C-H) and are much greater in magnitude than HFCs between atoms of different molecules: the H-H HFCs are found to be smaller by two orders of magnitude compared to C-C or C-H HFCs.

The largest HFCs are found at zero pair distance and describe the interactions of an atom with itself (self-HFCs: SHFCs). SHFCs are typically the largest in magnitude because they correspond to the (negative) sum of interactions an atom is exposed to from all the other ones in the supercell, when it is displaced from its equilibrium position. It is important to note that SHFCs have a different sign than the HFCs between different atoms. These SHFCs are calculated by *PHONOPY's* internal routines by applying the translation invariance symmetries (*acoustic sum rules*). The SHFCs are convenient measures to assess the contributions from interactions beyond the nearest neighbor: if the SHFCs are much smaller than the (negative) sum of nearest neighbor HFCs, this suggests relevant long-range interactions.

It is, however, important to note that the cartesian HFCs can only be directly correlated to frequencies in simple cases, since the calculation of frequencies includes several mathematical operations like the diagonalization of the entire dynamical matrix. To be able to relate HFCs to directional atomic motion, one would have to analyze the HFC-tensors component-wise, which is beyond the scope of this work. Therefore, the HFCs cannot be directly used to draw quantitative conclusions about specific vibrations.

In the following we will briefly comment on the methodological trends before we discuss their impact on the observed phonon frequencies. A more detailed comparison of the HFCs obtained with the different methodology shows that there are relatively large differences in the individual values (see Fig. S13). It is shown in Fig. S13(a,b) that all considered DFTB-based approaches as well as MOF-FF yield comparable values of the C-C (S)HFCs, while, interestingly, the COMPASS FF and the GAFF show distinct discrepancies: (i) COMPASS overestimates the magnitudes of the C-C nearest neighbor interactions (i.e., they have large negative values for that force field). The fact that all nearest neighbor C-C HFCs in Fig. S13(c) are too large leads to the observed particularly large self-HFCs in Fig. S13(a). (ii) GAFF leads to an even more severe overestimation of the nearest neighbor C-C HFCs than COMPASS, giving rise to massively overestimated SHFCs.

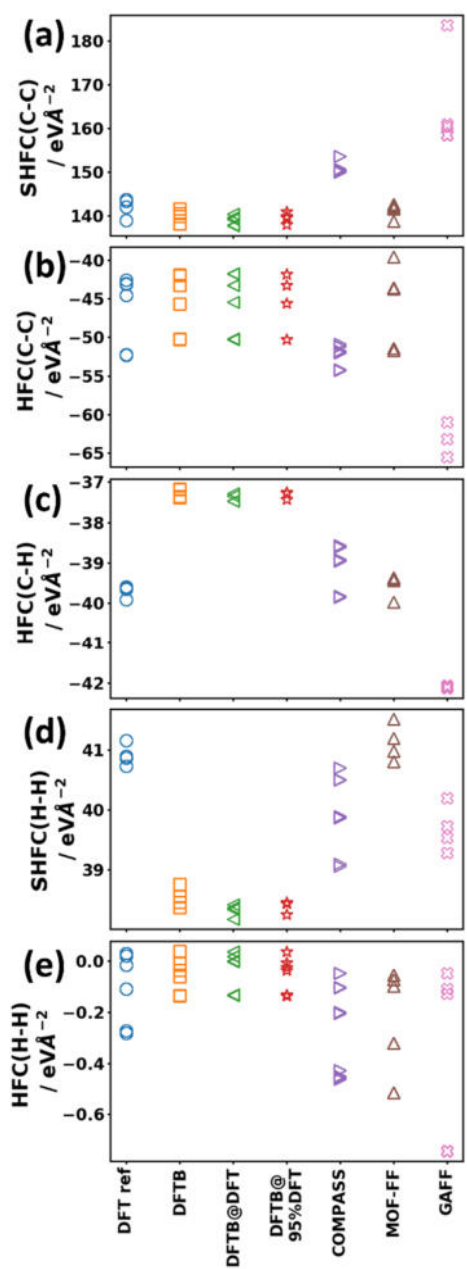

**Fig. S13:** *Harmonic force constants (HFCs) – i.e. traces of the rank 2 force constant tensors – obtained from different levels of theory. (a,d) Self-HFCs for carbon and hydrogen atoms, respectively. (b,c,e) HFCs for nearest neighbor interactions sorted according to the involved atomic species (C-C, C-H, H-H).*

Furthermore, Fig. S13(c-e) show that DFTB-based approaches consistently underestimate the magnitude of HFCs involving hydrogen, explaining the observation that the frequencies of C-H stretching vibrations are significantly underestimated. MOF-FF is able to reproduce the magnitude of the C-H HFCs. Relative the DFT reference data, GAFF yields discrepancies of similar magnitude but as DFTB (albeit with a different sign). Interestingly, in all cases except

26 / 31

for the GAFF results, the H-H SHFCs approximately equal the negative sum of the shown nearest neighbor C-H [Fig. S13(c)] and H-H [Fig. S13(e)] interactions. GAFF, however, in both cases overestimates the respective HFCs, but still shows underestimated H-H SHFCs, which must be the (negative) sum of C-H and H-H HFCs. This finding is a strong indicator for relevant interaction beyond the pair distances of covalently bonded atoms. Indeed, HFCs can be found for the next-nearest C-H interactions (~1.8…2.3 Å distance), which differ by the reference HFCs at these distances by fa factor of ~-3.6 (see Fig. S14).

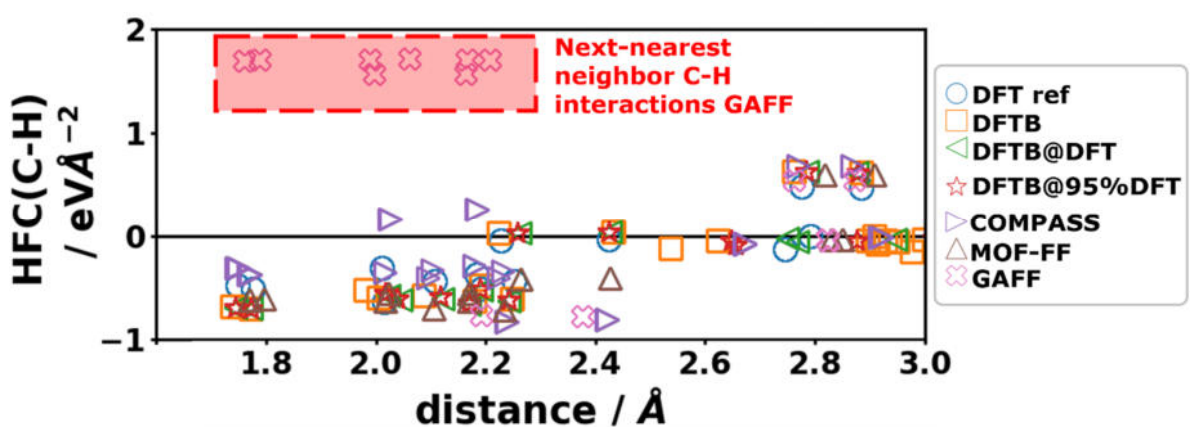

**Fig. S14:** *Trace of harmonic force constant tensors (HFCs) of C-H interactions as a function of distance of the involved atoms (considering periodic boundary conditions). The significantly overestimated HFCs from the GAFF are highlighted with a red box.*

The interpretation of the influence of the HFCs on the C-H stretching vibrations is slightly more involved. In the case of the DFTB-based approaches the interpretation is relatively straightforward: the too small C-H HFCs and H-H (S)HFCs are clear indicators why C-H stretching frequencies are underestimated that much. MOF-FF (COMPASS) shows slight tendencies to overestimate (underestimate) the corresponding frequencies, which is in agreement with the trends from the respective (S)HFCs. However, the GAFF results seem to be contradictive at first glance. The C-H HFCs are massively overestimated, while the H-H SFCs as well as the C-H stretching frequencies are too small. Therefore, in the case of GAFF, the influence of the underestimated H-H SFCs (due to the wrong next-nearest neighbor C-H HFCs; see Fig. S14) outweighs the effect of the overestimated C-H HFCs for the C-H stretching vibrations. The C-H bending frequencies are, however, overestimated by more than 10 THz. This observation is again in agreement with the too large C-H HFCs. This case is a good example, that it is difficult and often impossible to draw direct conclusions from the HFCs for all but the simplest cases.

## 14. Differences in thermodynamic properties at higher temperatures

In the main manuscript, we show heat capacities up to temperatures of 400 K. At that temperature the heat capacity of naphthalene is still far from saturating at the classical limit. Because of the high-frequency C-H stretching vibrations (above ~90 THz), the envelope function $f_C$ (see main text) must broaden to a large extent – i.e. by going to high temperatures – to reach those modes. For that reason, the heat capacity only approaches its classical limit ($3N\,k_B$) at temperatures beyond ~3500 K (see Fig. S15).

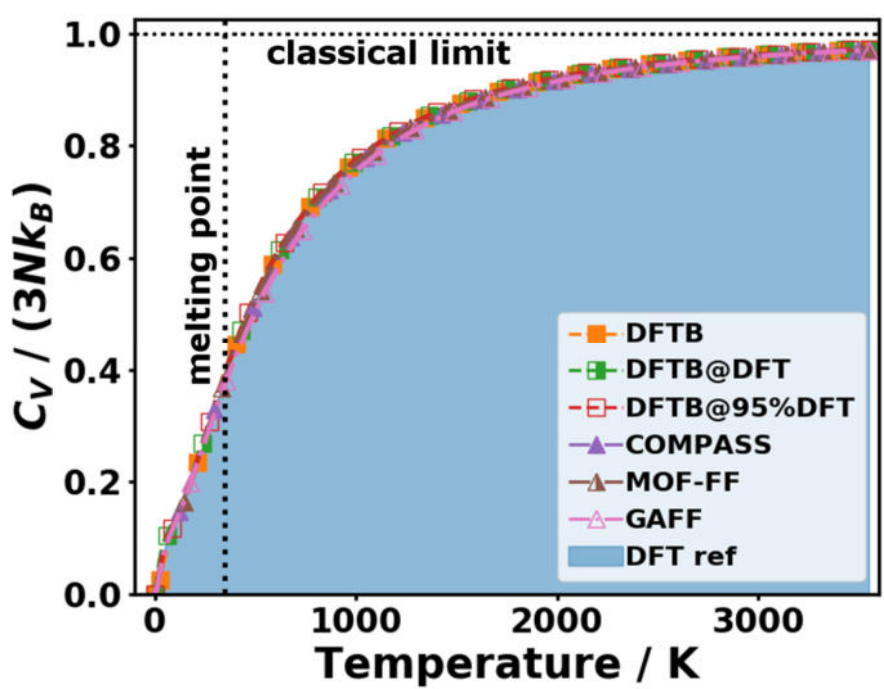

**Fig. S15:** *Saturation behavior of the phonon heat capacity $C_V$ as a function of temperature.*

Although the melting point of crystalline naphthalene (at atmospheric pressure) is ~353 K [19], it is still instructive to hypothetically track the errors in thermal properties to higher temperatures. Fig. 16 shows the difference in (normalized) heat capacities with respect to the DFT/D3-BJ reference calculation. As the heat capacity is per definition a quantity that saturates at high temperatures (Dulong-Petit limit), the deviations also converge to zero. The fastest convergence is obviously reached by MOF-FF and COMPASS for the reasons discussed in the main text.

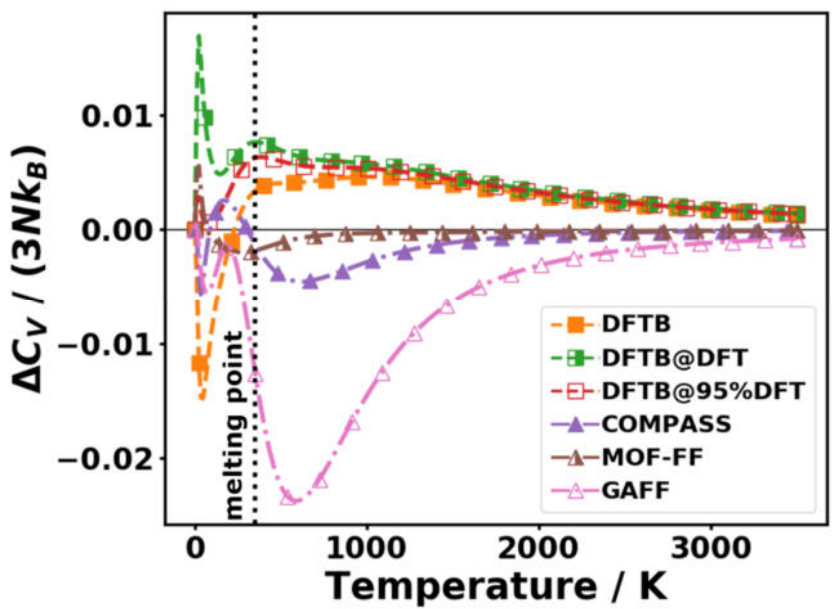

**Fig. S16:** *Difference in $C_V$ with respect to the DFT reference (DFT ref) as a function of temperature shown for very large temperatures.*

The free energy, however, does not approach a saturation value. Therefore, the differences increase for large T (T →∞). Again for MOF-FF and COMPASS, the error at 3500 K is only ~±0.3 eV suggesting a relatively robust description of the free energy over a wide temperature range.

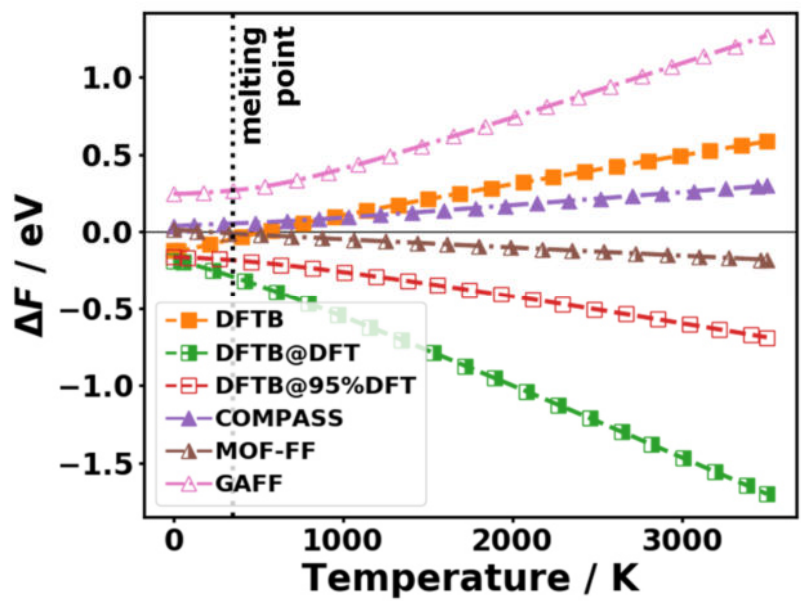

**Fig. S17:** *Difference in free energy F with respect to the reference as a function of temperature in a wide temperature range.*